\documentclass[12pt]{article}
\usepackage{amsmath,amssymb,graphicx,color,leqno,hyperref,url,float}

\usepackage[a4paper, total={6.5in, 8.75in}]{geometry}

\usepackage{enumitem}
\usepackage{lscape}

\newcommand{\mbf}[1]{\mbox{\boldmath $#1$}}
\newcommand{\ba}{{\mbf \beta}}

\newtheorem{cor}{Corollary}[section]
\newtheorem{lem}{Lemma}[section]
\newtheorem{rem}{Remark}[section]
\newtheorem{thm}{Theorem}[section]

\newtheorem{Def}{Definition}[section]

\numberwithin{thm}{section}

\numberwithin{equation}{section}

\newcommand{\cA}{{\cal A}}
\newcommand{\cB}{{\cal B}}

\newcommand{\cM}{{\cal M}}
\newcommand{\cN}{{\cal N}}

\def\ba{\begin{array}}
\def\bc{\begin{center}}
\def\bd{\begin{description}}
\def\be{\begin{enumerate}}
\def\ea{\end{array}}
\def\ec{\end{center}}
\def\ed{\end{description}}
\def\edt{\end{document}}
\def\ee{\end{enumerate}}
\def\ben{\begin{equation}}
\def\benn{\begin{equation*}}
\def\een{\end{equation}}
\def\eenn{\end{equation*}}
\def\benr{\begin{eqnarray}}
\def\eenr{\end{eqnarray}}
\def\benrr{\begin{eqnarray*}}
\def\eenrr{\end{eqnarray*}}

\def\al{\alpha}
\def\b{\beta}

\def\del{\delta}
\def\edt{\end{document}}

\def\g{\gamma}
\def\G{\Gamma}
\def\h{\hat}
\def\ka{\kappa}

\def\hs{\hskip}
\def\iny{\infty}
\def\ka{\kappa}

\def\la{\lambda}
\def\lel{\label}

\def\mb{\mbox}
\def\noi{\noindent}

\def\nn{\nonumber}

\def\Om{\Omega}

\def\r{\ref}

\def\ro{\rho}

\def\si{\sigma}

\def\Si{\Sigma}
\def\t{\tau}

\def\vep{\varepsilon}

\def\vs{\vskip}

\def\R{{\mathbb R}}

\begin{document}

\def\tecr{\textcolor{red}}
\def\tecb{\textcolor{blue}}
\def\so{\sout}

\def\tecb{\textcolor[rgb]{0.00,0.00,1.00}}
\bc
{\Large{\bf Two Stage Non-penalized Corrected Least Squares for High Dimensional Linear Models with Measurement error or Missing Covariates}}
\\[0.2cm]
Abhishek Kaul\footnote{Corresponding author. E-mail: abhishek.kaul@nih.gov, Research supported in part by the Intramural Research Program of the NIH, National Institute of Environmental Health Sciences (Z01 ES101744-
04)}, Hira L. Koul\footnote{Research supported in part by the NSF DMS grant 1205271}, Akshita Chawla and Soumendra N. Lahiri\footnote{Research supported in part by the NSF DMS grant 1310068.}\\[.1cm]

NIEHS, Michigan State University, Merck Research Laboratories and North Carolina State University

\ec
\vs .1in
{\renewcommand{\baselinestretch}{1}
\begin{abstract}
{\renewcommand{\baselinestretch}{1}
This paper provides an alternative to penalized estimators for
estimation and variable selection in high dimensional
linear regression models with measurement error or missing
covariates. We propose estimation via bias corrected
least squares after model selection. We show that by separating
model selection and estimation, it is
possible to achieve an improved rate of convergence of
the $\ell_2$ estimation error compared to the rate $\sqrt{s\log p/n}$
achieved by simultaneous estimation and variable selection
methods such as $\ell_1$ penalized corrected least squares.
If the correct model is selected with high probability then the $\ell_2$
rate of convergence for the proposed method is indeed the oracle rate of $\sqrt{s/n}.$ Here $s,$
$p$ are the number of non zero parameters and the model dimension,
respectively, and $n$ is the sample size.
Under very general model selection criteria, the proposed
method is  computationally simpler and statistically at
least as efficient as the $\ell_1$ penalized corrected least squares method,
performs model selection without the availability of the  bias correction matrix, and
is able to provide estimates with only a small sub-block of the bias correction covariance matrix of
order $s\times s$ in comparison to the $p\times p$  correction
matrix required for computation of the $\ell_1$ penalized version.
Furthermore we show that the model selection requirements are met by a correlation
screening type method and the $\ell_1$ penalized corrected least squares method.
Also, the proposed methodology when applied to the estimation
of precision matrices with missing observations, is seen to
perform at least
as well as existing $\ell_1$ penalty based methods. All results are supported empirically
by a simulation study. }
\end{abstract} }
\noi {\bf Keywords:} High Dimension, Measurement Error, Missing Data.

\section{Introduction}
Linear regression models with noisy or missing covariates are abound in variety of
scientific fields including econometrics, epidemiology and finance. Particular examples
of such data include the human microbiome expression data measuring relative
abundances of bacteria in the human body, which is
often observed only partially, i.e., with several missing observations
and gene expression data that are often corrupted with noise or
missing values. It is well known that ignoring this measurement error or missing-ness
leads to biased parameter estimates, see, e.g., Carroll, Ruppert, Stefansky and
Crainiceanu (2006) and Fuller (1987).

In the high dimensional setting where the number of parameters may vastly exceed the
sample size, several authors including Liang and Li (2009),  Loh and Wainwright (2012),
S{\o}rensen, Thoresen and Frigessi (2014), and
Kaul and Koul (2015), have studied estimators for these models.
The common thread of these papers being minimization of an appropriate bias corrected
loss function penalized by the $\ell_1$ norm of
the parameter vector of interest. This approach provides consistent estimates that are
also computationally efficient. However, defining the bias corrected loss function in
fact requires a bias correction matrix which is typically estimated from data.
This matrix being itself high dimensional makes its estimation and thus the
implementation of existing methods challenging, if not infeasible.

In this paper we propose a two step estimator for these models and analyse its efficiency
in model selection
and the rate of $\ell_2$ error in estimation. 
By separating model selection and estimation, it is possible to improve upon the rate of
$\ell_2$ error in estimation, 
compared to $\ell_1$ penalized methods.
Furthermore, our methodology 
requires only a small sub-block of the bias correction
matrix. Thus providing more accurate estimates with lesser information input in
comparison to $\ell_1$ penalized methods.
The main reason for this being that $\ell_1$ penalized methods are biased by construction.

Loh and Wainwright (2012) show that $\ell_1$ penalized corrected least squares method
achieves the rate $\sqrt{s\log
p/n}$ of the $\ell_2$ estimation error, under appropriate conditions. They also empirically
show that this rate is optimal. Here $p$ is the dimension of the parameter vector, $s$
represents the number of non zero mean parameters in the model and $n$ is the sample size.  In
comparison, our two stage
methodology enjoys three major advantages. First, the possibility of performing model selection
without the availability of the bias correction matrix. Second, being able to provide estimates
with only a small sub-block of the bias correction matrix. Lastly, provided one has a
reasonable control on the number of incorrectly identified regressors $(\h m),$ i.e., provided
$\h m =O_P(s),$  the proposed method performs at least as well as $\ell_1$- penalized methods.
In addition, if the correct model is selected from the first step with probability (w.p.)
converging to $1,$ then the rate of convergence of the $\ell_2$-error for the proposed method
is shown to be indeed the optimal rate of $\sqrt{s/n}.$
We also apply the methodology developed
to the problem of precision matrix estimation with observations corrupted with missing values and
similarly show that the estimates thus obtained are more efficient in comparison to its $\ell_1$ penalized
counterpart.

To the best of our knowledge, such two stage refitting procedures were first introduced by
Candes and Tao (2007) in the context of Dantzig selector for high dimensional classical linear
regression where $X$ is fully observed, and have been investigated  by
Belloni and Chernuzhokov (2013)  with least squares loss again in the linear
regression setup without measurement error. In particular, the latter provide a rigorous analysis of the rate of convergence
of the $\ell_2$ error for the two stage refitting procedure.

Finally, we perform a series of simulated experiments to confirm our theoretical findings. We
show empirically  that in addition to having higher efficiency in estimation, our methodology provides more accurate model identification compared to the $\ell_1$ penalized counterpart and is also computationally faster for larger data sets.

The rest of this paper is organized as follows. Section 2 describes the model under consideration and
introduces the notation required for the analysis. Section 3 describes the first step
model selection procedure and investigates some theoretical properties of the
two possible methods, which can be used to achieve this goal consistently. Section 4 provides some
 theoretical properties of the second step estimation procedure and describes the associated rates of
convergence of estimation error. We then provide an algorithm for precision matrix estimation with
observations corrupted with missing data. Section 5 provides a series of simulated experiments.
All proofs are relegated to the appendix.

\section{Model Setup}

We begin by describing the models under consideration. 
Let $x_i=(x_{i1},\cdots,x_{ip})'$, $i=1,\cdots, n,$ be  vectors of random design
variables, where for any vector $a$, $a'$ denotes its transpose. Let $y_i$'s denote the
responses, which are related to $x_i$'s by the relations
\benr\lel{rm}
&& y_{i}= x_i'\beta_0 + \vep_i, \quad \mb{for some $\b_{0}=(\b_{01},...,\b_{0p})'\in \R^p$},  \,\, 1\le i\le n.
\eenr
Here $\b_0$ is the parameter vector of interest, and
$\vep=(\vep_1,...,\vep_n)'$ is an $n-$dimensional vector whose components are
i.i.d.\,\,Gaussian random variables (r.v.'s) with variance $\si_{\vep}^2,$ i.e.,
$\vep_i\sim_{i.i.d}\cN(0,\si_{\vep}^2),$
$1\le i\le n.$ Furthermore, the design variables $x_i$'s are not observed directly.
Instead, we observe surrogates $z_i$, $1\le i\le n,$ obeying one of the
following two models.
\vs .1cm
{\bf Additive noise:}
\benr\lel{me}
z_i=x_i+w_i,\quad  1\le i\le n.
\eenr
The covariate noise vectors $w_i=(w_{i1},..,w_{ip})',$ $1\le i \le n,$ are
assumed to be
i.i.d.\,\,r.v.'s. Furthermore, $w_i,$ $x_i,$ and $\vep_i,$ $1\le i\le n$, are assumed
to be mutually independent.
\vs .1cm

{\bf Missing covariates:}
\benr\lel{miss}
z_i=x_i\oplus w_i,\quad  1\le i\le n.
\eenr
Here $\oplus$ represents componentwise product and $w_i=(w_{i1},...,w_{ip})',$ with
the components $\{w_{ij}, 1\le i\le n\}\sim_{i.i.d.}\mb{Bernoulli}(1-\rho_j),$
$1\le j\le p$.
\vs .2cm
Let $X=(x_1,...,x_n)'$ be the unobserved $n\times p$ design matrix and similarly
define the $n\times p$ matrices $Z,$ $W$ with the corresponding vectors. For the case of
additive noise, the random matrices $X$ and
$W$ are assumed to be sub-Gaussian as defined by Loh and Wainwright (2012). This
definition is restated below for the convenience of the reader.
\begin{Def}{\bf (sub-Gaussian matrices)}\lel{subg} {\rm We say that a random matrix $X\in
\R^{n\times p}$ is sub-Gaussian with parameters $(\Sigma_x,\si_x^2)$ if the following two
conditions hold.
\begin{enumerate}
\item Each row $x_i'\in\R^{p}$ of $X$ is sampled independently from a zero-mean
    distribution with covariance $\Si_x,$ $1\le i
\le n$.
\item For any unit vector $\delta\in\R^{p}$ the random variable
$\delta'x_i$ is sub-Gaussian in the usual univariate sense with parameter at most
$\si_x^2,$ $1\le i\le n.$
\end{enumerate}}
\end{Def}

\begin{rem}\lel{zsubg}{\rm
An elementary property of sub-Gaussianity and $X$ and $W$ being sub-Gaussian imply that in the case of additive noise,
 $Z$ is also sub-Gaussian. Also, Loh and Wainwright (2012) show as part of the proof
of Lemma 4 of their supplement that for the case of missing covariates, the random matrix $Z$ is
also sub-Gaussian with parameter $\si_x^2,$ i.e.,
with the same parameter as for the unobserved sub-Gaussian
random matrix $X.$}
\end{rem}


\section{Notation, Assumptions and Conventions}
The parameters  $p$ and $s$ are assumed to
diverge with the sample size $n,$ however this dependence is suppressed for clarity of
the exposition. For the same reason we do not exhibit the dependence of the arrays of
$x_i$'s and $z_i$'s on $n$.
For any vector $\delta\in\R^{p},$ define the support of $\delta$ as
$\mb{supp}(\del)=\big\{j\in\{1,2,...,p\}; \del_j\ne 0.\big\}.$
The $\ell_2$ norm of $\del$ is denoted by $\|\cdot\|_2$ and $|\delta|$ shall denote the
componentwise absolute value vector. For any two collection of indices $S$ and
$\tilde S,$ we represent
$\tilde S-S$ as the collection of indices in $\tilde S$ but not in $S$. The
cardinality of an index
set $S$ will be denoted by either  $\mb{card}(S)$, and
$\|\delta\|_0:=\mb{card}\big(\mb{supp}(\delta)\big).$ For any two sequences
$\{a_n\}$ and $\{b_n\}$ of real numbers,
$a_n\preceq b_n$ means that for some constant $0<c_0<\iny$,
 $a_n\le c_0 b_n$, for $n$ large enough.
Similarly, $a_n\preceq_P b_n$ shall denote that $a_n\preceq b_n$ in
probability.
For matrices $M_1$ and $M_2$ we denote $M_1\oplus M_2$ and $M_1\ominus M_2$ as the
component wise product and division, respectively.
For a subset $A\subseteq \{1,2,\cdots,p\}$, $b_A$ denote the vector of components of $b$
with indices in $A$. Also, all limits are taken as $n\to\iny,$ unless mentioned otherwise.
Lastly, $0<c_0,c_1,c_2<\iny,$ and
$0<c_3<1$ shall denote generic constants that may be different in different contexts.

In the above setup we shall consider the model (\r{rm}) in the high dimensional setting
where the dimension $p$ of  $\b_0$ is allowed to grow
exponentially with $n.$ In addition $\b_0$ is assumed to be sparse, i.e., only a small proportion of
the  parameters are assumed to be non zero. In the sequel,
\benrr\lel{T}
T=\mb{supp}(\b_0), \qquad \mb{$T^c$ denote its compliment set.}
\eenrr
By definition, $\mb{card}(T)=s.$

Decompose $\b_0=\big(\b_{0T}',\b_{0T^c}'\big)'$ into its non zero and zero components,
and similarly partition  $n\times p$ matrices $X$ and $Z$ into columns corresponding to the indices of $\b_0$, i.e.,
$
X=\big(X_T,X_{T^c}\big),\,  Z= \big(Z_T,Z_{T^c}\big).
$
Also a $p\times p$ matrix $\Si$ is partitioned as
\benr\lel{sigpar}
\Si = \left(\begin{matrix} \Si_{TT} & \Si_{TT^c}\\ \Si_{T^cT} & \Si_{T^cT^c}
\end{matrix}\right).
\eenr
Throughout, the parameters $s,$ $p$ and $n$ are assumed to satisfy
\benrr 
s\log p/n =o(1).
\eenrr

\vs .15cm
Define
\benr\lel{smg}
\h\g^{\mb{add}}=n^{-1}Z'y\qquad\mb{and}\qquad \h\g^{\mb{miss}}=n^{-1}Z'y\ominus (\bf
{1}-{\boldsymbol\rho}),
\eenr
where {\bf 1} is a $p$-dimensional vector of ones and ${\mbf \rho}=(\rho_1,..,\rho_p)'.$
The entities $\h\g^{\mb{add}}$ and $\h\g^{\mb{miss}}$
serve as measures of correlation between $X$ and $y$ in the additive error and missing
covariates cases, respectively. Also define,
\benr
\G^{\mb{add}}=n^{-1}Z'Z-\Sigma_w,\quad\mb{and}\quad \G^{\mb{miss}}=n^{-1}Z'Z\ominus M,
\eenr
for the additive error and missing covariate cases, respectively. Here, $M=\big[M_{ij}\big]_{i,j=1,...,p}$ is a
$p\times p$ matrix with
\benr\lel{M}
M_{i,j}=\begin{cases}
      (1-\rho_i)(1-\rho_j) \,\,; & i\ne j \\
      (1-\rho_i)           \,\,; & i=j.  \end{cases}
\eenr

Next, we state the needed assumptions.


\vspace{2mm}
\noi{\large {\bf Assumptions:}}
\vspace{1mm}

\noi {\bf (A1)} {\bf Additive errors: } In the model (\r{me}), the measurement error
matrix $W=(w_1,...,w_n)'$ is assumed to be sub-Gaussian as defined in (\r{subg}) and $w_i,$ $x_i$ and
$\vep_i$ are assumed to be mutually independent for all $1\le i\le n.$\\[.15cm]
{\bf (A2)} {\bf Missing covariates}: The components of the vector $w_i$ in the model (\r{miss}) 
are such that $\{w_{ij}, 1\le i\le n\}$ are i.i.d.\,\,Bernoulli$(1-\rho_j),
1\le j\le p.$ Also assume that
$0\le\rho_{\max}:=\max\{\rho_j\,;\, 1\le j\le p\}<1.$
Furthermore $w_i$'s are
mutually independent of $x_i$ and $\vep_i,$ for all $1\le i\le n$.

\vs .2cm
\noi {\bf Unobserved design variables $X$:}\\[.1cm]
\noi{\bf (A3)} Assume that the covariance matrix of $X$ satisfies the following conditions, where part (i) is for additive errors and part (ii) is for missing covariates, and where $\ro_{max}$ is as in (A2).
\benrr
&& \mb{(i)} \quad \max |\Si_{T^cT}^x\b_T^0|+2\frac{\si_z}{c_0}(\si_{\vep}+\si_x\|\b_0\|_2)\sqrt{\frac{c_1\log p}{n}}< \min |\Si_{TT}^x\b_T^0|.\\
&& \mb{(ii)} \quad \max |\Si_{T^cT}^x\b_T^0|+2\frac{\si_x}{c_0(1-\rho_{\max})}(\si_{\vep}+\si_x\|\b_0\|_2)
\sqrt{\frac{c_1\log p}{n}} \\
&&\hs 3in < \min |\Si_{TT}^x\b_T^0|.
\eenrr
This assumption is similar to Condition F of Genovese et al.\,\,(2012) and is also
reminiscent of the `faithfulness condition' of B\"uhlmann, Kalisch and Maathuis (2009).
In the noiseless setting, it 
is necessary and sufficient for exact recovery
of the support of $\b_0,$ (Thm.\,\,2, Genovese et al.\,\,2012).

\vspace{2mm}
\noi{\bf Random matrices $\Gamma^{\mb{add}}$ and $\Gamma^{\mb{miss}}$:} \\[.1cm]
\noi {\bf RE:} A matrix $\Gamma$  is said to satisfy the lower restricted eigenvalue condition
with curvature $\al_1>0$ and tolerance $\tau>0$ if
\benr
\delta'\Gamma\delta\ge \al_1\|\delta\|_2^2-\t\|\delta\|_1^2,\qquad \mb{for all}\,\,\delta\in\R^p.
\eenr
{\bf RSE$(k_n)$:} For any $m\le k_n,$ a matrix $\Gamma$ is said to satisfy a lower and upper restricted
sparse eigenvalue condition with constants $\kappa(m),\phi(m)>0$, respectively, if
\benr\lel{rse}
&& \kappa(m):= \inf_{\|\delta_{T^c}\|_0\le m,\,\delta\ne 0}\frac{\delta'\Gamma\delta}{\|\delta\|_2^2}>0, \quad
 \phi(m):= \sup_{\|\delta_{T^c}\|_0\le m,\,\delta\ne 0
}\frac{\delta'\Gamma\delta}{\|\delta\|_2^2}<\iny. 
\eenr

Assumption {\bf RE} was introduced by Loh and Wainwright (2012). They prove that this condition holds
for $\G^{\mb{add}}$ and $\G^{\mb{miss}}$, with asymptotic probability $1$ with appropriate choices of
$\al_1$ and $\tau.$

Assumption {\bf RSE} controls the minimum and maximum eigenvalues of certain sub-blocks
of the matrix $\Gamma.$ Lemma \r{r1} below shows that this condition
is satisfied by the matrices $\G^{\mb{add}}$ and $\G^{\mb{miss}}$ with asymptotic probability $1$.

\vs .2cm
\noi{\bf (A4)} {\bf Parameter vector $\b_0$:} The minimum magnitude of the components of
$\b_0$ satisfies $\min_{j\in T}|\b_{0j}|\succeq \|\b_0\|_2\,s\log p/n.$

\vs .1cm
The following lemma shows that the random matrices $\G^{\mb{add}}$ and
$\G^{\mb{miss}}$ satisfy the condition {\bf RSE} with asymptotic probability $1,$ under suitable assumptions. Let
\benrr\lel{xeigen}
\kappa_x(m):= \inf_{\|\delta_{T^c}\|_0\le m,\,\delta\ne 0 }\frac{\delta'\Si_x\delta}{\|\delta\|_2^2},
\qquad  \phi_x(m):= \sup_{\|\delta_{T^c}\|_0\le m,\,\delta\ne 0 }\frac{\delta'\Si_x\delta}{\|\delta\|_2^2}.
\eenrr

\begin{lem}\lel{r1}{\bf (Plausibility of RSE).} Let $k_n$ be any positive sequence satisfying
$k_n \log p= o(n).$ Suppose condition {\bf (A1)} for the additive error model or condition {\bf (A2)} for the
missing covariate model hold. Also assume that some constants $\ka_x$ and $\phi_x$,
\benr\lel{xeigen}
0<\ka_x\le\ka_x(m)\le \phi_x(m)\le \phi_x<\iny, \quad \mb{for all $m\le k_n$}.
\eenr
Then, with $\G=\G^{\mb{add}}$ or $\G=\G^{\mb{miss}},$ the following conditions
\benr
&& \kappa(m):= \inf_{\|\delta_{T^c}\|_0\le m,\,\delta\ne
0}\frac{\delta'\Gamma\delta}{\|\delta\|_2^2}>0,\quad
\phi(m):= \sup_{\|\delta_{T^c}\|_0\le m,\,\delta\ne 0
}\frac{\delta'\Gamma\delta}{\|\delta\|_2^2}<\iny,\nn
\eenr
hold uniformly over any $m\le k_n$ with $\ka(m)\ge \ka_x/2$ and $\phi(m)\le 2\phi_x$, w.p.  at least $1-2c_3\exp(-s)/(1-1/e)$,  for all  sufficiently large $n$.
\end{lem}

This lemma shows that for any positive sequence $k_n$ satisfying $k_n\log p=o(n),$
condition {\bf RSE$(k_n)$}
is satisfied by $\G^{\mb{add}}$ and $\G^{\mb{miss}},$ with the lower and
upper restricted eigenvalues $\ka(m)$ and $\phi(m)$ being bounded below and
above, respectively, for large $n,$ with high probability. This lemma shall play a useful
role in the development of the methodology to follow.

\section{Step 1: Model Selection}
The objective of this first step is to recover the support $T$ of the parameter vector $\b_0$ from the
observed variables $Z$ and $y.$ In the sequel $\h T$ denotes the estimate of the
support $T$ of $\b_0$ given by the model selection procedure and $\h m$ denotes
the number of noise variables selected, i.e.,
\benrr
\h m=\mb {card}(\h T-T).
\eenrr
We propose the following two possible methods for selecting $\h T$.
\vs .15cm
\noi {\bf CS} Screen the corrected absolute correlation vector $\big|\h\g^{\mb{add}}\big|$
or $\big|\h\g^{\mb{miss}}\big|$ to select a certain number of indices that are largest in magnitude.
The intuition behind this is the same as that of the sure independence screening proposed by Fan and Lv (2008).
To see this equivalence for the additive error case notice that $Z'y=X'y +W'y.$ Now,
by assumption $W$ is independent of $y,$ thus the correlation structure of $n^{-1}Z'y$
will asymptotically be the same as that of $n^{-1}X'y.$ These ideas are made rigorous
below.\\[.15cm]
$\ell_1$-{\bf CLS} Use $\ell_1$ penalized bias corrected least squares as proposed by Loh and
Wainwright (2012) to select the indices of the non zero estimates. This estimator is defined as
\benr\lel{l1cls}
\h \b=\operatornamewithlimits{arg\,min}_{\|\b\|_1\le b_0\sqrt{s}} \Big\{\h Q_n(\b)+\la_n\|\b\|_1\Big\},
\quad\qquad \la_n>0.
\eenr
where $b_0$ is a suitably chosen constant and
\benr\lel{qnb}
\h Q_n  (\b):=\frac{1}{2}\b'\Gamma\b-\h\g'\b,
\eenr
where $\G$ and $\h\g$ are
chosen as the corresponding versions in the additive errors or the missing covariates cases. The selected model is $\h T=\mb{supp}(\h\b).$

\vs .2cm

We begin with the analysis of the {\bf CS} method. Consider the absolute value of the correlation vector
$|\h\g| :=(|\h\g_1|,...,|\h\g_p|)'$ defined in (\r{smg}) between the observed variable $Z$ and
$y,$ and
let $r(\h \g)=(r_1(|\h\g|),...,r_p(|\h\g|))'$ denote the vector of descending
ranks of the components of the vector $\h\g,$ where rank one signifies the
highest magnitude. Then the {\bf CS} method estimates the set of non zero indices
by
\benr\lel{htan}
&& \h T (a_n)= \h T(a_n, Z,y, p)=\{j\,;\, r_j(\h\g)\le a_n, \,\, 1\le j\le p\},
\eenr
where $a_n$ is a known sequence of positive numbers such that $a_n/s\le c,$ for some constant $c\ge 1.$
The following theorem shows that this procedure identifies the support
of the parameter vector along with providing a reasonable control on the false positives.

\begin{thm}\lel{mst1}
If either conditions {\bf(A1)} and {\bf (A3i)} hold for the case of
additive errors (\r{me}) or conditions {\bf(A2)} and {\bf (A3ii)}
hold for the case of missing covariates (\r{miss}), then the estimated
set of non zero indices $\h T(a_n)$ of (\r{htan}) satisfies the following.
\benrr
\mb{(i)}\,\,\,\, T\subseteq \h T(a_n),\qquad \mb{(ii)}\,\,\,\,\h m \preceq s,
\eenrr
w.p. at least $1-c_1\exp(-c_2\log p),$ for all sufficiently large $n$.
\end{thm}


\begin{rem}{\rm A closer look at the proof of Theorem \r{mst1} shows that if the
cardinality of the set $T$ is known and we let $a_n\to s$
in (\r{htan}), then $P\big(\h T=T\big)\to 1.$ }
\end{rem}

In view of Theorem \r{mst1}, choosing $a_n$ appropriately leads to identification of the
support of the parameter vector along with a control on the false positives. However, the
choice of this thresholding level $a_n$ is determined by the number of non-zero
components $s,$ which in practice is unknown. Thus, as is the case with $\ell_1$
penalized methods, we shall treat $a_n$ as a tuning parameter and provide a data based
strategy to optimally choose this parameter in Section 6.

As stated earlier, the implementation of this method does not require the
knowledge of the matrix $\Si_w$ or $M$. This is especially useful in the
case of additive errors where $\Si_w$ is unknown, since we by-pass estimating a $p$
dimensional $\Si_w$ from a very low number or typically available replicates of $Z.$ In addition, this method comes
at a cheap computational cost.

Next, we proceed to the $\ell_1-{\bf CLS}$ method for model selection.
Before proceeding, a point of caution
here is that this method is not useful for the case of additive errors due to the
unavailability of $\Si_w$. On the other hand, for the case of
missing covariates we can estimate $\rho_j$ for all $1\le j\le p$ by the empirical
average of the number of observed entries per column of $Z.$ This in turn enables us
to estimate the matrix $M$ and to implement  $\ell_1$-penalized bias
corrected least squares in this case. Thus, the analysis to follow shall focus on model
selection by  $\ell_1$-{\bf CLS} method only for the case of missing covariates.

Another technical reason for not  using $\ell_1$-{\bf CLS} in the case of additive errors is the non convexity of the
loss function $\h Q_n(\b).$ In comparison, $\h Q_n(\b)$ is convex in the case of missing
covariates, which plays a key role proving the desired model selection property of this methodology.

We begin with the following additional assumption. For some $r>0$,
\benr\lel{ldc}
\|\G^{\mb{miss}}\b_0-\h\g^{\mb{miss}}\|_{\iny}\le r\|\b_0\|_2\sqrt{\log p}/\sqrt{n}.
\eenr
Then we have the following model selection result.

\begin{thm}\lel{mst2}
In addition to (\r{rm}), (\r{miss}), and (A4), suppose the conditions
lower-{\bf RE} and (\r{ldc}) hold  for $\G^{\mb{miss}}$ and $\h\g^{\mb{miss}}.$
Then for the method $\ell_1$-{\bf CLS} with $\la_n\ge 4r \|\b_0\|_2\sqrt{\log p}/\sqrt{n}$ in
(\r{l1cls}), where $r$ is as in (\r{ldc}), we have, with $\h T=\mb{supp}(\h\b)$,
\benr
\mb{(i)}\,\,\,\, T\subseteq \h T, \hs 1in\mb{(ii)}\,\,\,\, \sqrt{\h m}\le c_0\phi(\h m)\sqrt{s}/\al_1.\nn
\eenr
\end{thm}

\begin{rem}\lel{rms2}{\rm
The result of Theorem \r{mst2} is not accompanied by a probabilistic statement since this result follows by deterministic arguments on the event where the required assumptions hold. In addition Loh and
Wainwright (2012) (Theorem 1 and Corollary 2) show that that the conditions lower-{\bf RE} and (\r{ldc}) hold for $\G^{\mb{miss}}$ and the pair $(\G^{\mb{miss}},\h\g^{\mb{miss}}),$
respectively, w.p. at least $1-c_1\exp(-c_2\log p)$, with
\benr
\al_1=\la_{\min}(\Si_x)/2\quad \mb{and}\quad   r=c_0\frac{\si_x}{1-\rho_{\max}}\big(\si_{\vep}+\frac{\si_x}{1-\rho_{\max}}\big),\nn
\eenr
where $\la_{\min}(\Si_x)$ represents the minimum eigenvalue of the matrix $\Si_x.$}
\end{rem}

\begin{rem}{\rm Recall that for the case of missing covariates, $\h Q_n(\b)$ is convex,
and hence, by standard results via first order optimality conditions, $\h m\le n,$ see, e.g., Lemma 5
of Tibshirani (2013). Thus Theorem \r{mst2} immediately implies that with high
probability, $\h m\preceq \phi(n)s.$ However, this bound is not sharp since $\phi(n)$ may
diverge with $n.$ From here, following the strategy of Belloni and Chernozhukov (2013),
we extend the result to obtain the bound $\h m\preceq s$ under an additional
assumption.  This will be implied by the following lemma.}
\end{rem}
\begin{lem}\lel{slre2} Under the conditions of Theorem \r{mst2},
\benr
\h m\le \frac{c_0}{\al_1}s\Big[\min_{m\in\cM}\phi(m\wedge n)\Big],
\eenr
where $\cM=\Big\{m\in {\bf N}\,;\, m>\frac{2c_0}{\al_1}s\phi\big(m\wedge n\big)\Big\}.$
\end{lem}

This lemma is a consequence of Theorem \r{mst2} and
Lemma 2 of Belloni and Chernozhukov (2013) and thus the short proof is omitted. For
details see, page 14 of Belloni and Chernozhukov (2013). As a consequence of this lemma,
under the additional assumption
\benr\lel{las}
\min_{m\in \cM}\phi(m\wedge n)\le c_0,  
\eenr
$\h m\preceq s.$ This result, together with Theorem \r{mst2} and
Remark \r{rms2}, yields that for the case of missing covariates, the model selected via
$\ell_1-{\bf CLS}$ satisfies
\benr\lel{mselprop}
  T\subseteq \h T,\qquad\quad \h m\preceq s,
\eenr
w.p. at least $1-c_1\exp(-c_2\log p)$, for all sufficiently large $n$.
This concludes this section on the recovery of the support of $\b_0.$ We now proceed to
the estimation of $\b_0.$

\section{Step 2: Estimation}
This section shall investigate the estimation properties of the following estimator. With $\h Q_n(\cdotp)$ as in (\r{qnb}),  define the post selection corrected least squares estimator of $\b_0$ as
\benr\lel{tilb}
\tilde \b=\operatornamewithlimits{arg\,min}_{\b\in\R^p} \h Q_n(\b);\,\,\, \b_j=0,\quad
\mb{for each}\,\, j\in\h T^c.
\eenr
We shall show that the estimator $\tilde \b$ performs at least as well as $\ell_1$ penalized
corrected least squares in terms of the rate of convergence of $\ell_2$ estimation error,
under suitable assumptions on model selection.  More interestingly, $\tilde \b$ has the
potential to outperform $\ell_1$ penalized methods, depending on the first step model
selection. In fact ${\tilde\b}$ attains the oracle rate $\sqrt{s/n}$ under perfect
model selection $(w.p.  \to 1).$ Furthermore, the implementation of the proposed
estimator requires the knowledge of only a sub-block \big($O(s)$-dimensional\big)
of the bias correction matrices $\Si_w$ or $M$ in the additive error or missing
covariate cases, respectively.

For any constant $0< c_3\le 1$ and a universal constant $D,$ let
\benr\lel{enm}
&& e_n(m,c_3)=\sqrt{\frac{m \log p}{n}} +\sqrt{\frac{(m+s)\log (D)}{n}}
+\sqrt{\frac{m+s+\log (1/c_3)}{n}}.
\eenr

Consider the following assumption.
\benr\lel{dc}
\sup_{\|\delta_{T^{c}}\|_0\le
m,\,;\,\|\del\|_2>0}\frac{1}{\|\delta\|_2}\Big|\delta'\G\b_0-\delta'\h\g\Big|
\le c_0 r\|\b_0\|_2e_{n}(m,c_3).
\eenr
Here 
$0<r<\iny$ is a suitably chosen constant
depending on the two sources of noise $W$ and $\vep.$  Later in this section we show that this uniform
bound holds with asymptotic probability $1$ for both pairs $\big(\G^{\mb{add}}, \h\g^{\mb{add}}\big)$ and $\big(\G^{\mb{miss}}, \h\g^{\mb{miss}}\big).$ We now state
the main result of this section.

\begin{thm}\lel{t2} Suppose model is selected by the {\bf CS} method and assumptions of Theorem \r{mst1}
hold. Furthermore, assume that the pairs $(\Gamma^{\mb{add}}, \h\g^{\mb{add}})$ for the additive error case or $(\Gamma^{\mb{miss}}, \h\g^{\mb{miss}})$ for the missing covariate case satisfy
the uniform deviation condition in (\r{dc}) and the condition lower-{\bf RSE}$(a_n)$ with $a_n$ as in (\r{htan}). Then there exists a universal
positive constant $c_0$ such that
\benr
\|\tilde\b-\b_0\|_2\le \frac{1}{\ka(\h m)}c_0r\|\b_0\|_2e_n(\h m,c_3),
\eenr
holds, w.p. at least $1-c_1\exp(-c_2\log p)-\big(6c_3\exp(-s)\big/(1-1/e)\big),$ for all sufficiently large $n$.

\end{thm}



\begin{cor}\lel{cor1}
Suppose the conditions of Theorem \r{t2} hold, and that $\|\b_0\|_2\le b_0,$
for some constant $b_0<\iny.$ Then
\benr
\|\tilde \b-\b_0\|_2\preceq_P\begin{cases}
      \sqrt{\frac{s \log p}{n}} \,\,; & \mb{in general} \\
      \sqrt{\frac{s}{n}}+\sqrt{\frac{o(1)s\log p}{n}}\,\,; &\mb{if}\,\, a_n/s\to 1^{+},\\
     \sqrt{\frac{s}{n}}\,\,; & \mb{if}\,\, a_n=s.
       \end{cases}
\eenr
\end{cor}

The proof of this corollary is a direct consequence of Theorem \r{t2} and is thus
omitted. An immediate consequence of this corollary is that implementing the two
stage corrected least squares with the first stage model selection done via
the {\bf CS} method  will result in estimates that perform at least as well as
$\ell_1$ penalized counterparts.  More importantly, the two stage method has room for
improvement for the rate of convergence. In contrast, $\ell_1$ penalized methods have a
rate of $\sqrt{s\log p/n}$ which is empirically known to be optimal, see, e.g.
Loh and Wainwright (2012). In fact under perfect $(w.p. \to 1)$ model selection,
$\tilde \b$ achieves the $\sqrt{s/n},$ which is the oracle rate of convergence.

The second useful aspect of this method is that implementing
the second step estimation requires only an $O(s)$ dimensional block of the $p$
dimensional bias correction matrix $\Si_w$ or $M$ to be known or estimated. In
comparison,  the $\ell_1$ penalized method for simultaneous model selection and
estimation requires the entire $p$ dimensional matrix. Keeping in mind that the
dimension $p$ can be growing exponentially with $n$, estimating  $\Si_w$
from the low number of typically available replicates of the design variables may be
infeasible.

Next, we focus on the case of missing covariates where model selection is done via
$\ell_1$-{\bf CLS} method and estimation via (\r{tilb}). This shall again
yield estimates that are at least as efficient as the estimates based on
$\ell_1-{\bf CLS}$ method and shall allow room for improvement in its efficiency.

\begin{thm}\lel{t3} Suppose the model (\r{rm}) and (\r{miss}) holds, and let
model selection be done via $\ell_1$-{\bf CLS} method. Assume conditions of
Theorem \r{mst2} and in addition assume that the pair $\big(\Gamma^{\mb{miss}}, \h\g^{\mb{miss}}\big)$ satisfies the uniform deviation condition
(\r{dc}) and the matrix $\G^{\mb{miss}}$ satisfies condition lower-{\bf RSE}$(b_n)$ and (\r{las}) for any $b_n=O(s).$  Then there exists a
universal positive constant $c_0$ such that
\benrr
\|\tilde\b-\b_0\|_2\le \frac{1}{\ka(\h m)}c_0r\|\b_0\|_2e_n(\h m,c_3).
\eenrr
In particular, if $\|\b_0\|_2\preceq 1$, then 
the rate
of convergence described here is at least
\benrr
\|\tilde\b-\b_0\|_2\preceq \sqrt{\frac{s\log p}{n}}.
\eenrr
\end{thm}


The results of Theorem \r{t3} are not accompanied by a probabilistic statement since this result follows by deterministic arguments on the event where the required assumptions hold. In view of Lemma \r{r1} and Lemma
\r{udcc}, all assumptions made on random quantities made here hold with asymptotic probability $1.$ Thus, the conclusions of this theorem hold with asymptotic probability $1.$

The proof of Theorem \r{t3} essentially uses the property (\r{mselprop})
from the first step model selection and is similar to that of Theorem \r{t2}, hence omitted.
This theorem may also be extended easily to any model selection procedure
satisfying (\r{mselprop}).

The only remaining part is to now show that the uniform deviation assumption (\r{dc}) holds with high probability. This forms the content of the following lemma.

\begin{lem}\lel{udcc} Suppose the model (\r{rm}) holds and that
the covariate noise $W$ satisfies conditions {\bf (A1)} and {\bf(A3)} for additive error
and missing covariate cases, respectively. Let
\benrr
r=\begin{cases}
      \si_z(\si_w+\si_{\vep})\hs 0.6in\,; & \mb{\rm for additive error} \\
      \frac{\si_x}{1-\rho_{\max}}\big(\si_{\vep}+\frac{\si_x}{1-\rho_{\max}}\big)\,\,; & \mb{\rm for missing covariates}.
       \end{cases}
\eenrr
Then, with $\G=\G^{add},\, \h \g=\h \g^{add}$ or $\G=\G^{miss}, \, \h \g=\h \g^{miss}$,
\benrr
\sup_{\|\delta_{T^{c}}\|_0\le
m\,;\,\|\del\|_2>0}\frac{1}{\|\delta\|_2}\Big|\delta'\G\b_0-\delta'\h\g\Big|\le c_0
r\|\b_0\|_2e_{n}(m,c_3),
\eenrr
w.p. at least $1-6c_3 e^{-s}/(1-1/e),$ for all sufficiently large $n$.
\end{lem}

\subsection{Application to estimation of precision matrices with missing observations}

Estimation of covariance and precision matrices plays an important role in several
statistical analyses including the principal component analysis,
linear/quadratic discriminant analysis, and graphical modeling. This problem has
been extensively researched in the case where the observations are i.i.d.\,\,vectors from
a multivariate sub-Gaussian distribution, see, 
Meinhausen and B\"uhlmann (2006), Friedman, Hastie and Tibshirani (2007) and
Bickel and Levina (2008). On the other hand, when the observed vector is corrupted
by missing variables, Loh and Wainwirght (2012) propose an algorithm based on the
$\ell_1$ penalized corrected least squares estimator, which provides consistent
estimates.

Suppose observations $X_i\in\R^p,$ $1\le i\le n$ are i.i.d.\,\,$\cN(0,\Si),$ where $\Si$
is a positive definite matrix. Then it is well known, see, e.g.,  Anderson (2003), that
for each $1\le j\le p,$
the conditional distribution of the $j^{th}$ component $X_i^j$, given the rest $X_i^{-j},$
is again normal distribution, i.e.,
\benr
X_i^{j}\Big|X_{i}^{-j}\sim\cN\Big(\Si_{j,-j}\Si_{-j,-j}^{-1}X_i^{-j}\,\,,\,\,
\Si_{jj}-\Si_{j,-j}\Si_{-j,-j}^{-1}\Si_{-j,j}\Big).\nn
\eenr
This result can equivalently be written as the linear relation,
\benr\lel{msg}
X_i^{j}=X_i^{-jT}\theta^{j}+\vep^{j},\quad 1\le i\le n,
\eenr
where $\theta^{j}=\Si_{-j,-j}^{-1}\Si_{-j,j}$ is a $p-1$ dimensional vector and
$\vep^{j}$ is a vector of i.i.d. Gaussian r.v.'s, independent of $X^{-j}.$ Here
$\Si_{-j,-j}$ represents the sub-block of $\Si$ with the $j^{th}$ row and column removed.
The precision matrix $\Theta:=\Si^{-1}$ can then be reconstructed from $\theta^{j},$
$1\le j\le p$ as follows,
\benr
&& \Theta_{jj}=d_j,\,\,\,\mb{and}\,\,\,\Theta_{-j,j}=-d_j\theta^{j},\quad\mb{where}\quad
d_j:=(\Si_{jj}-\Si_{j,-j}\theta^{j})^{-1}.
\eenr

When $Z_i=X_i\oplus W_i$ is observed in place of $X_i,$ $1\le i\le n$ with missing
observations as described in (\r{miss}), we can use the methodology proposed in sections 4
and 5 above to estimate the parameters $\theta^{j}$ of the model (\r{msg}) for every $1\le j\le
p.$ Note that the response and predictor variables both have missing observations in this
case, however the proofs of our results in the previous sections can be easily seen to
hold under this setup.

For any matrix  $A=(a_{ij})_{1\le i,j\le p}$, let $\|A\|_2=\max_{1\le j\le p}\big(\sum_{i=1}^{p}
a_{ij}^2\big)^{1/2}.$ Also define $\la_{\min},$ $\la_{\max}$ as the minimum and maximum
eigenvalues of the covariance matrix of interest $\Si.$ Then we have the following
algorithm.

\vspace{2mm}
\noi{\bf Algorithm 1:}
\vs .2cm
\noi 1. Let $\h\Si=n^{-1}Z'Z\ominus M$ with $M$ as defined in (\r{M}). For each
    $1\le j\le p,$ define
\benr
(\G^{j},\h\g^{j})=\Big(\h\Si_{-j,-j},n^{-1}Z^{-jT}Z^{j}\ominus(
{\mbf 1}-{\mbf \rho^{-j}})(1- {\mbf \rho}_j)\Big),\nn
\eenr
and estimate the support of $\theta^j$  by the {\bf CS} method for model selection, i.e.,
with the thresholding level $a_n=c_0s,$ $c_0>1,$ let
\benr
\h T_j (a_n)= \{k\,;\, r_k(\h\g)\le a_n, \,\, 1\le k\le (p-1)\}.\nn
\eenr
2. Obtain estimates $\h\theta^{j}$ by the following optimization,
\benr\lel{gest}
\h \theta^j=\operatornamewithlimits{arg\,min}_{\|\theta\|_1\le b_0\sqrt{s}} \h Q^j(\theta)\,;\, \theta_k=0,\qquad \mb{for each}\,\, k\in\h T_j^c.
\eenr

\noi 3. Substitute $\h\Si$ and $\h\theta^{j},$ to obtain $\h d_j=(\h\Si_{jj}-\h\Si_{j,-j}\h\theta^{j})^{-1}$ and complete the estimated precision matrix $\tilde\Theta$ with $\tilde\Theta_{-j,j}=-\h d_j\h\theta^{j}$ and $\tilde\Theta_{j,j}=\h d_j$
\vs .15cm
\noi 4. Set $\h\Theta=\mb{arg min}_{\Theta\in S^{p}}\|\Theta-\tilde\Theta\|_2,$ where $S^{p}$ is the collection of symmetric matrices.
\vs .2cm

Note that we have placed an additional restriction on the parameter space in step 2 of the algorithm, i.e., $\|\theta^j\|_1\le b_0\sqrt{s},$ where $s$ represents $\mb{card}(T_j).$ This additional restriction does not influence the proofs of Section 4 and 5.

The choice of the thresholding level $a_n=c_0s$ is required for our proofs.
However in practice $a_n$ is a data based tuning parameter as described earlier, see also Remark \r{tune}.

We now proceed to providing consistency in estimation of the above algorithm. The
assumptions required for this purpose are restated below in the present context.

\vspace{2mm}
\noi {\bf Assumptions:}
\vs.2cm
\noi{\bf (G1)} The vectors $\theta^{j}=\Si_{-j,-j}^{-1}\Si_{-j,j}$ are $s$-sparse, i.e., for all $1\le j\le p,$ $|T_j|\le s,$ where $T_j=\mb{Supp}(\theta^{j}).$ Furthermore, assume that $\|\theta^j\|_1\le b_0\sqrt{s},$ $1\le j\le p$ for some constant $b_0<\iny.$
\vs .2cm
\noi{\bf (G2)} The covariance matrix $\Si$ has bounded maximum and minimum eigenvalues, i.e., $0<\la_{\min}(\Si)\le\la_{\max}(\Si)<\iny,$ and satisfies the following relation for all $1\le j\le p,$
\benrr
&&\max|\Si_{-j,-j}^{T_j^cT_j}\theta_{T_j}^{j}|
+2\frac{\si}{c_0(1-\rho_{\max})^2}(\si_{\vep}
+\si_x\|\theta^j\|_2)\sqrt{\frac{c_1\log p}{n}}  \\
&& \hs .25in < \min |\Si_{-j,-j}^{T_jT_j}\theta_{T_j}^{j}|.
\eenrr
Here $\Si_{-j,-j}^{T_jT_j},$ and $\Si_{-j,-j}^{T_j^cT_j}$ are a partitions of $\Si_{-j,-j}$ as described in (\r{sigpar}).
\vs .2cm
\noi{\bf (G3)} The pairs $(\G^j,\h\g^j), 1\le j\le p,$ satisfy the following uniform deviation condition.
For each $1\le j\le p,$ and for some constant $r<\iny.$
\benr
 \sup_{\|\delta_{T_j^{c}}\|_0\le m,\,;\,\|\del\|_2>0}\frac{1}{\|\delta\|_2}\Big|\delta'\G^j\theta^j-\delta'\h\g^j\Big|\le c_0 r\|\theta^j\|_2e_{n}(m,c_3).
\eenr
\begin{thm}\lel{gmc} In addition to conditions {\bf(G1)}, {\bf (G2)} and {\bf (G3)}, assume that
the lower-{\bf RSE}$(a_n)$ condition holds uniformly over the matrices $\G^j$, and
\benr\lel{covb}
\big\|\h\Si-\Si\big\|_{\iny}\le c_0\si_{x}\sqrt{\frac{\log p}{n}}.
\eenr
Then the estimated precision matrix $\h\Theta$ provided by Algorithm 1 satisfies
\benr
\big\|\h\Theta-\Theta\big\|_2\le c_0\Big(C_2^2+\frac{C_1^2}{\la_{\min}^2}+\frac{\la_{\max}}{\la_{\min}}C_2\Big)^{1/2}e_n(a_n-s, c_3),\nn
\eenr
w.p. converging to 1, where
$$C_1=r\la_{\max}/\la_{min}, \qquad C_2=\Big(b_0\si_x+ \big(r\la_{\max}^2/\la_{\min}^2\big)\Big)\Big/\la_{\min}^2.
$$
\end{thm}

\begin{rem} {\rm
Assumption (\r{covb}) is a standard assumption for high dimensional covariance
recovery. It can be shown to hold with asymptotic probability $1$ with
an appropriate choice of constant $c_0,$ see Yuan (2010). The uniform bound of
Assumption {\bf (G3)} can be shown to hold using the same arguments as in Lemma \r{udcc}.
Lastly, the condition lower-{\bf RSE} can be shown to hold uniformly over
$1\le j\le p$ with high probability by applying arguments of Lemma \r{r1} along with
the observation that in this case uniformly over all $1\le j\le p,$
\benr
0<\la_{\min}\le k_{{-j,-j}}(m)\le \phi_{{-j,-j}}(m)\le \la_{\max}<\iny,
\eenr
for any $1\le m\le k_n.$ Here $\ka_{-j,-j}$ and $\phi_{-j,-j}$ are as defined in
(\r{xeigen}) with $\Si_x$ replaced by $\Si_{-j,-j}.$}
\end{rem}

\section{Simulation Study}

In this section we numerically analyse the performance of the methodology developed in
this paper. We implement our two step methodology with model selection done via method
{\bf (CS)} and refer to these as the post selection estimates. In the following we shall
compare the post selection estimates with the $\ell_1$ penalized corrected least squares
estimates of Loh and Wainwright (L\&W) and the ordinary Lasso which disregards covariate
noise or missing-ness.

\begin{rem}\lel{tune} {\rm
{\it Tuning Parameter:} The model selection step in our post selection estimates involves
choosing an appropriate value of the thresholding level $a_n.$ To choose this tuning
parameter, we employ standard cross validation logic, i.e., the  {\bf CS} method
is used to
select models for a grid of values of $a_n\in \{1,2,..,\min\{p,(n/\log p)\}\}$ and
corresponding estimates $\tilde\b_{a_n}$ are obtained via (\r{tilb}). These estimates are
then used on  an independent test set to compute the corrected least squares loss
\benrr
L(\tilde\b_{a_n})=\frac{1}{2}\tilde\b_{a_n}'\G\tilde\b_{a_n}-\h\g'\tilde\b_{a_n}.
\eenrr
The tuning parameter $a_n$ is  chosen as a minimizer of this criteria.
Naturally, larger the grid chosen for $a_n,$ more is the computation time necessary. To
maintain fairness of comparisons the same cross validation approach is used to choose the
tuning parameter $\la$ of $\ell_1$ penalized corrected least squares estimator. The
tuning parameter for the ordinary Lasso is chosen via similar cross validation with the
loss function chosen as ordinary least squares. In both cases $\la$ is allowed to range
from zero to one with increments of 0.05.}
\end{rem}
All simulations are performed in R, the estimates of ordinary Lasso are obtained
using
the package {\it glmnet} developed by Friedman et al.\,\,(2013). Post selection estimates
$\tilde\b$ and L\&W estimates are obtained via the projected gradient descent algorithm,
see, e.g. Agarwal, Neghban and Wainwright (2012) and Loh and Wainwright (2012).

\subsection{Simulation Setup and Results}
We begin with the regression setting with additive error or missing covariates. Here, the
unobserved design variables $\{x_{ij},\,1\le i\le n,\,1\le j\le p,\}$ are chosen as
i.i.d.\,\,r.v.'s from a standard normal distribution. Components of the parameter
vector $\b_0$ are generated independently from a uniform distribution with  support
$(-4,-1)\cup(1,4).$ The
model errors $\vep_i,$ $1\le i\le n$ are generated as i.i.d.\,\,$\cN(0,\si_{\vep}^2)$ with
$\si_{\vep}=0.25.$

To evaluate performance of the estimators, two cases each are presented for additive errors
and missing covariates. An independent data set for every combination of the following
$(n,p,s)$ settings is generated,\\[.1cm]
$\bullet$ \,\,  Sample size $n$ ranging from $100$ to $500$ with increments of $40.$
Model dimension $p$ ranging from $100$ to $500$ with increments of $65$ and the number of non zero parameters $s=4,8.$ Thus leading to 154($11\times 7\times 2$) independent models.\\[.1cm]
$\bullet$ \,\, $p=750,$ $s=4$ and $n$ ranging from $50$ to $500$ with increments of $5,$ thus leading to 91 independent models.
\vs .1cm
The three estimators to be compared are computed for each generated model and we report the following measures for comparison, (1) relative estimation error,  REE:=$\|\b-\b_0\|_2/\|\b_0\|_2,$ (2) number of false positives, i.e., number of incorrectly identified zero components, and (3) computation time (in seconds) required to obtain estimates.

\vspace{2mm}
\noi
{{\bf Example 1. Regression Setting:} Suppose the model (\r{rm}) holds and consider the following two cases.\\[.1cm]
$\bullet$ {\it Additive Error}: The covariate noise variables $W_i,\,1\le i\le n$ are assumed to be i.i.d.\,\,Gaussian with mean zero and covariance matrix $\Si_w=c_w\big[\si_{ij}^w\big]_{i,j=1,..,p}$, where $\si_{ij}^w=0.5^{|i-j|}$ and $c_w=0.25.$ Simulation results are illustrated in Figure \r{regsim}: {\bf Ad1-Ad4}.\\[.1cm]
$\bullet$ {\it Missing Covariates}: The missing covariates are generated as described in assumption {\bf (A2)} where $\rho_j,$ $1\le j\le p$ are chosen independently from a uniform distribution over the support $(0.05,0.75)$. Simulation results are illustrated in Figure \r{regsim}: {\bf M1-M4}.

\begin{figure}[H]
\begin{minipage}{.48\textwidth}
{\bc\bf Ad1\ec}
\includegraphics[scale=0.30]{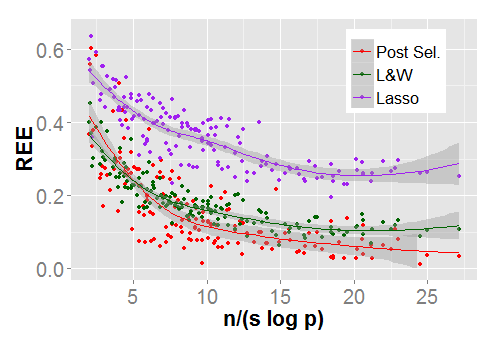} 
\end{minipage}
\begin{minipage}{.48\textwidth}
{\bc\bf Ad2\ec}
\includegraphics[scale=0.30]{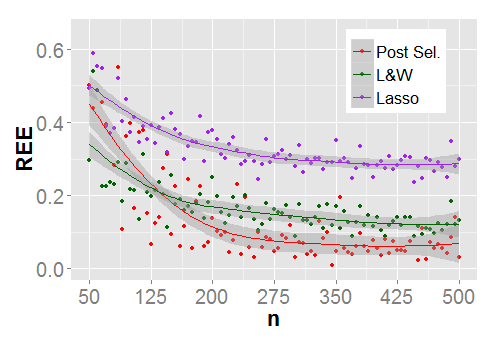} 
\end{minipage}
\end{figure}
\vspace{-5mm}
\begin{figure}[H]
\begin{minipage}{.48\textwidth}
{\bc\bf Ad3\ec}
\includegraphics[scale=0.30]{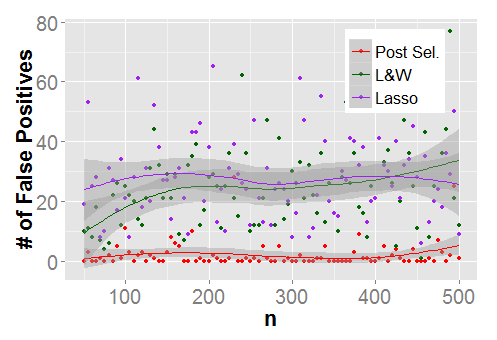} 
\end{minipage}
\begin{minipage}{.48\textwidth}
{\bc\bf Ad4\ec}
\includegraphics[scale=0.30]{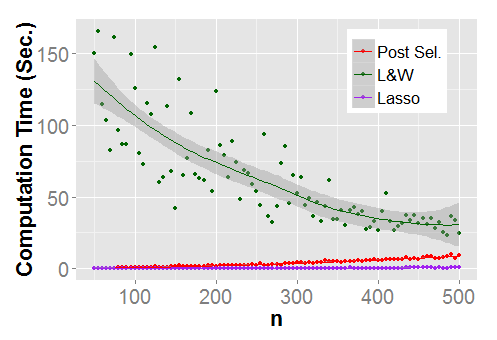} 
\end{minipage}
\end{figure}
\vspace{-5mm}
\begin{figure}[H]
\begin{minipage}{.48\textwidth}
{\bc\bf M1\ec}
\includegraphics[scale=0.30]{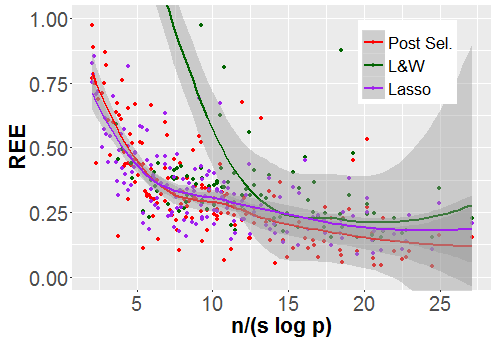} 
\end{minipage}
\begin{minipage}{.48\textwidth}
{\bc\bf M2\ec}
\includegraphics[scale=0.30]{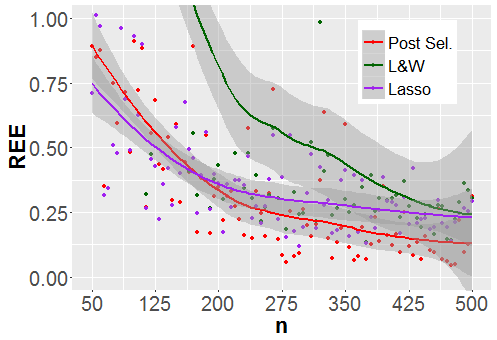} 
\end{minipage}
\end{figure}
\vspace{-5mm}
\begin{figure}[H]
\begin{minipage}{.48\textwidth}
{\bc\bf M3\ec}
\includegraphics[scale=0.30]{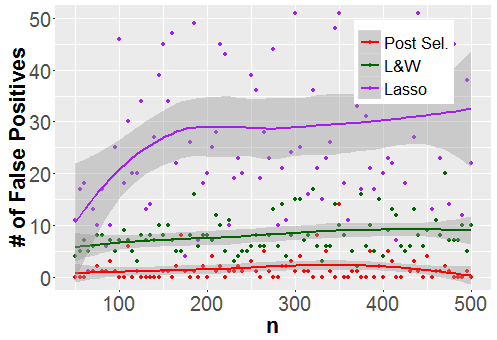} 
\end{minipage}%
\begin{minipage}{.48\textwidth}
\centering
{\bc\bf M4\ec}
\includegraphics[scale=0.30]{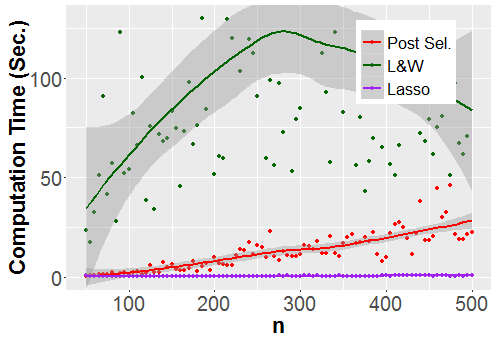} 
\end{minipage}
\caption{\footnotesize{ {\bf Ad1 \& M1} plots REE against $n/s\log p,$ here $n,p\in[100,500],$ and $s\in\{4,8\}$ for the additive error and missing covariate cases respectively. {\bf Ad2, Ad3, Ad4 \& M2, M3, M4}, plots REE, false positives, and computation time at p=750, against $n$ in the additive and missing cases respectively.}}
\lel{regsim}
\end{figure}

\noi $\bullet$ {\it Estimation accuracy}: The empirical results support the theoretical findings. Consistency in the $\ell_2$ estimation error of the post selection estimator is clearly observed. In addition, the post selection estimates nearly uniformly outperform the two other estimators, see Figure \r{regsim}: {\bf Ad1, Ad2, M1, M2}. The L\&W estimates perform marginally better at lower sample sizes for the additive error case, see Figure \r{regsim}: {\bf Ad1, Ad2}.\\[.1cm]
$\bullet$ {\it False positives}: In both additive and missing covariate cases, the post selection estimates are seen to provide a significant improvement in the control on false positives, see Figure \r{regsim}: {\bf Ad3, M3}. \\[.1cm]
$\bullet$ {\it Computation time}: The computation time for post selection estimates is significantly quicker in comparison to L\&W estimates and comparable to Lasso at larger values of $p$ and a fixed sample size $n.$ However
the computation time for post selection estimates increases with $n$ due to the increase in the grid size of $a_n$ for cross validation. In comparison, the computation time of L\&W and Lasso estimates decrease as the grid size for cross validation stays the same and numerical convergence becomes quicker with higher $n,$ see Figure \r{regsim}: {\bf Ad4, M4}.

\vs 0.1cm
\noi
{\bf Example 2. Graphical Models:}
In this example we examine the efficacy of the proposed algorithm in estimating the precision matrices for two types of Gaussian graphical models, namely band and cluster structured graphs. These precision matrices are generated by the package "fastclime" developed by Pang, Liu and Vanderbei (2014). For a $p$-dimensional graph, around $p/20$ band width or clusters are assumed in the two cases, respectively. The adjacency matrices of these graphs with $p=50$ are illustrated below
\begin{figure}[H]
\begin{minipage}{.48\textwidth}
\centering
\includegraphics[scale=0.25]{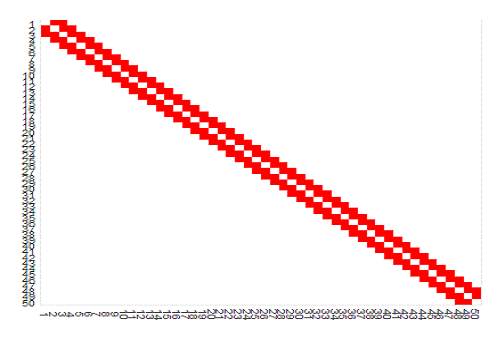} 
\end{minipage}
\begin{minipage}{.48\textwidth}
\centering
\includegraphics[scale=0.25]{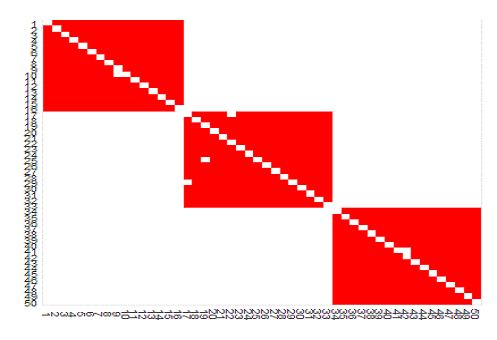} 
\end{minipage}
\caption{\footnotesize{Plots of adjacency matrices of banded and cluster precision matrices respectively.}}
\lel{omill}
\end{figure}
The precision matrices are generated so that the corresponding covariance matrix
$\Si=\Om^{-1}$ is normalized to have all diagonal components $1.$ For further details on
the construction of these matrices see, page 5 of Pang, Liu and Vanderbei (2014). Next,
the unobserved variables $X_i,$ $1\le i\le n,$ are generated as i.i.d. $\cN(0,c_x\Si)$
for $c_x=1,3.$ Missing-ness is then induced as $Z_i=X_i*W_i,$ $1\le i\le n$ in
accordance with (\r{miss}), where $\rho_j$, $1\le j \le p,$ are chosen independently from
a uniform distribution over the support $(0.05,0.75)$. For each model, we compute
estimates via the proposed Algorithm 1 and compare it to the estimates based on
the $\ell_1$ penalized version of L\&W. For performance comparison we report
$\|\h\Theta-\Theta\|_2,$ in addition to false positives identified in the matrix and
the computation time required to compute corresponding estimates.

\begin{figure}[H]
\begin{minipage}{.48\textwidth}
{\bc\bf CG1\ec}
\includegraphics[scale=0.30]{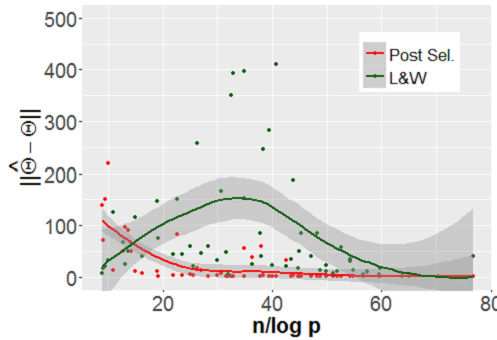} 
\end{minipage}
\begin{minipage}{.48\textwidth}
{\bc\bf CG2\ec}
\includegraphics[scale=0.30]{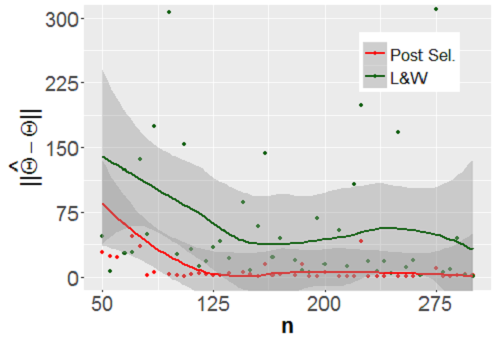} 
\end{minipage}
\end{figure}
\vspace{-5mm}
\begin{figure}[H]
\begin{minipage}{.48\textwidth}
{\bc\bf CG3\ec}
\includegraphics[scale=0.30]{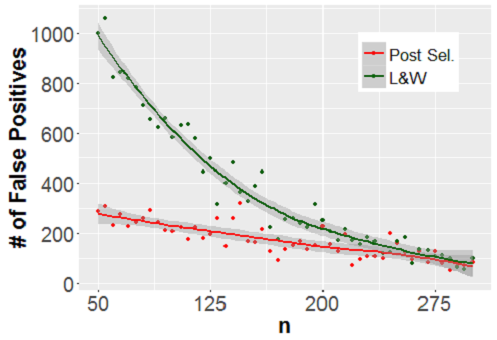} 
\end{minipage}%
\begin{minipage}{.48\textwidth}
\centering
{\bc\bf CG4\ec}
\includegraphics[scale=0.30]{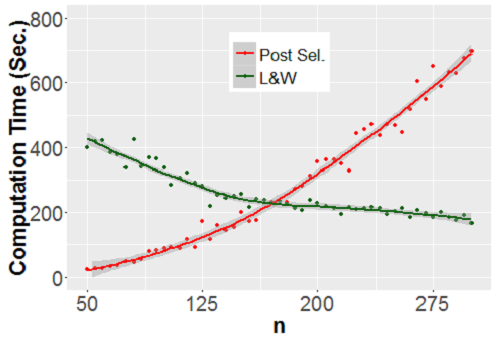} 
\end{minipage}
\end{figure}
\vspace{-5mm}
\begin{figure}[H]
\begin{minipage}{.48\textwidth}
{\bc\bf BG1\ec}
\includegraphics[scale=0.30]{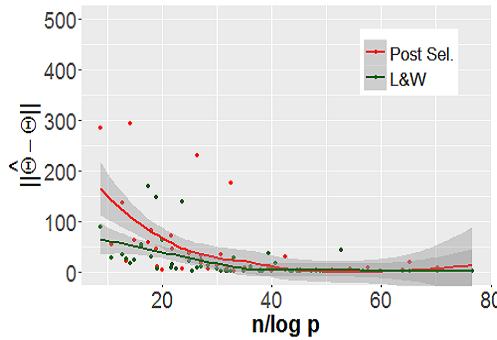} 
\end{minipage}
\begin{minipage}{.48\textwidth}
\centering
{\bc\bf BG2\ec}
\includegraphics[scale=0.30]{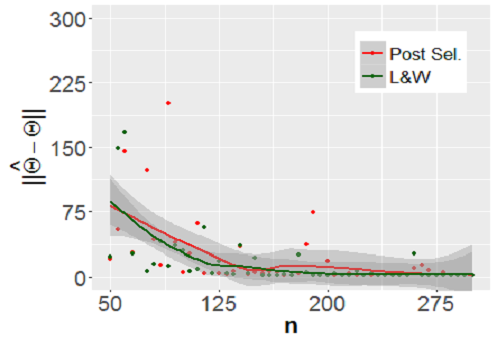} 
\end{minipage}
\end{figure}
\vspace{-5mm}
\begin{figure}[H]
\begin{minipage}{.48\textwidth}
\centering
{\bc\bf BG3\ec}
\includegraphics[scale=0.30]{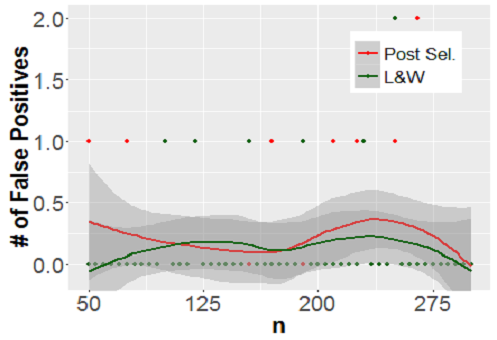} 
\end{minipage}
\begin{minipage}{.48\textwidth}
\centering
{\bc\bf BG4\ec}
\includegraphics[scale=0.30]{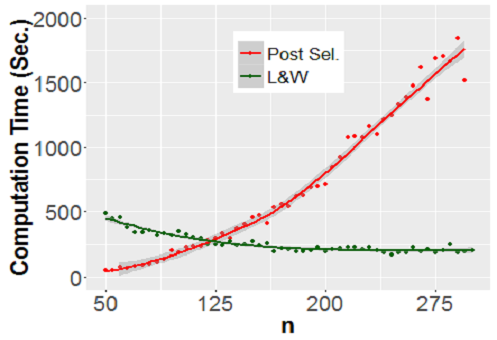} 
\end{minipage}
\caption{\footnotesize{ {\bf CG1 \& BG1} plots $\|\h\Theta-\Theta\|_2$ against $n/\log p,$ here $n, p\in[50,300],$ for the cluster and banded graph cases respectively. {\bf  CG2, CG3, CG4 \& BG2, BG3, BG4}, plots $\|\h\Theta-\Theta\|_2$, false positives, and computation time at p=100 against $n$ in the cluster and banded graph cases respectively. Also, $c_x=3,1$ in the cluster and banded graph cases respectively.}}
\lel{grafsim}
\end{figure}

\noi $\bullet$ {\it Estimation accuracy}\,: Post selection estimates provide consistent
estimates of $\Theta.$ In addition, they are uniformly superior in the case of the
cluster graph and perform about as well as L\&W estimates in the banded graph case,
see Figure \r{grafsim}: {\bf CG1, CG2 \& BG1, BG2}. This is 
due to the constant $c_x$ which is $3$ in the latter and $1$ in the banded graph case.
It is seen that the post selection estimates become uniformly superior as $c_x$ is
increased.\\[.1cm]
$\bullet$ {\it Computation time}\,: Although the computation time for L\&W estimates in
the settings presented here is significantly faster when $n$ increases, however it is
also observed that increasing the dimension $p$ significantly favors post selection
estimates in terms of computational efficiency.

\vspace{2mm}
\noi{\bf Note:} In Figures \r{regsim} and \r{grafsim}, three colors of each dot represent
a performance measure corresponding to an independently generated model for the three
estimates being compared. To measure the average performance over the independently
simulated models, non parametric regression lines and corresponding confidence bands are
drawn, these are made via the Loess method with its smoothing parameter set as $0.75$.


\section{Appendix}

\subsection{Proofs for Section 3}
The proofs to follow require a probability bound for centered sum of squares of
independent sub-Gaussian r.v.'s. This is facilitated by Lemma 14 of Loh and Wainwright
(2012) supplement which is restated below for completeness. This lemma in turn is a
direct corollary of Lemma 14 of Vershynin (2012).
\begin{lem}\lel{loh}
If $X\in\R^{n\times p_1}$ is any zero mean sub-Gaussian matrix with parameters
$(\Si_x,\si_x^2),$ then for any fixed unit vector $v\in\R^{p_1}$ and $t>0$,
\benr\lel{lw1}
&& P\left(\Big|\|Xv\|_2^2-E\|Xv\|_2^2\Big|\ge nt\right)\le 2
\exp\Big(-\frac{1}{c_0}n\min\big\{\frac{t^2}{\si_x^4},\frac{t}{\si_x^2}\big\}\Big),
\eenr
where $c_0>0$ is a universal constant. Moreover, if $Y\in\R^{n\times p_2}$ is a zero mean sub-Gaussian matrix with parameters $(\Si_y,\si^2_y)$ then for every $t>0$,
\benr\lel{lw2}
\lefteqn{
P\left(\Big\|n^{-1}Y'X-\mb{\rm Cov}(y_i,x_i)\Big\|_{\iny}\ge t\right) } \\
&\le& 6p_1p_2 \exp\Big(-c_0n\min \big\{\frac{t^2}{\si_x^2\si_y^2},\frac{t}{\si_x\si_y}\big\}\Big). \nn
\eenr
Here $y_i$ and $x_i$ represent the $i^{th}$ rows of $Y$ and $X$, respectively.
\end{lem}

To state the next lemma  we need to define
\benrr
&&\cB(m)=\{\del\in\R^p,\,;\, \|\del_{T^c}\|_0\le m\}, \quad \cB_1(m)=\{\delta\in\cB(m),\,;\,\|\del\|_2\le 1\}.
\eenrr
\begin{lem}\lel{l2} Let $X\in \R^{n\times p}$ be a sub-Gaussian matrix
with parameters $(\Sigma_x, \si^2_x).$ For a $c_3\in (0,1)$, let
$$
r_n=r_n(m,c_3)=\frac{\si^2_x}{3c_0} e_n(m,c_3),
$$
where $e_n(m,c_3)$ is as in (\r{enm}).
Then, for all sufficiently large $n$,
$$
P\left(\sup_{\delta\in \cB_1(m)}\Big|\frac{\|X\delta\|_2^2}{n}-E\frac{\|X\delta\|_2^2}{n}\Big|\ge r_n,\,\,\, {\mb{for all}}\,\,\, m\le n\right) \le  2c_3 e^{-s}/(1-1/e).
$$
\end{lem}

\vspace{3mm}
\noi {\bf Proof of Lemma \r{l2}.} For every $U\subseteq\{1,...,p\}$ with $\mb{card}(U-T)
\le m,$ define $S_U:=\{\delta\in\R^p;\, \|\delta\|_2\le 1;\, \textrm{supp}(\delta)
\subseteq U\}$ and note that $\cB_1(m)= \cup_{card(U-T)\le m}S_U.$ Let
$\cA=\{u_1,...,u_k\}$ be a $1/10$ cover of a fixed $S_U,$ i.e., for each $\delta\in S_U$
there exists $u_i\in\cA$ such that $\|\delta-u_i\|_2\le 1/10.$ It is known from Ledoux and
Talagrand (1991) or Loh and Wainwright (2012) (Supplementary materials, pg.17), that one
can construct $\cA$ such that $card(\cA)\le 100^{(m+s)}.$ Let $\psi(\delta_1,
\delta_2)=\delta_1'\big(\frac{X'X}{n}-\Si_x\big)\delta_2$. Then by
elementary algebra,
\benr\lel{l2eq1}
\psi(\delta,\delta)=\psi(u_i,u_i)+2\psi(\delta-u_i,u_i)+\psi(\delta-u_i,\delta-u_i)
\eenr
Note that by construction, there exists $u_i\in\cA$ such that
$10(\delta-u_i)\in S_U$ and thus $2\psi(\delta-u_i,u_i)=\frac{2}{10}\psi(\delta_1,u_i),$
for $\delta_1:=10(\delta-u_i)\in S_U.$

Now expressing the second term on the r.h.s. of (\r{l2eq1}) as,
\benrr
\frac{2}{10}\psi(\delta_1,u_i)&=&\frac{1}{10}\psi(\delta_1,\delta_1)+\frac{1}{10}\psi(u_i,u_i)-\frac{1}{10}\psi(\delta_1-u_i,\delta_1-u_i)\\
                    &\le& \frac{1}{10}\max_{i}|\psi(u_i,u_i)| + \frac{5}{10}\sup_{\delta\in S_U}|\psi(\delta,\delta)|,
\eenrr
where the last inequality follows since for any $\delta_1,\delta_2\in S_U$ we also have $\frac{1}{2}(\delta_1-\delta_2)\in S_U.$ Replacing this inequality in (\r{l2eq1}) we obtain
\benrr
&&\sup_{\delta\in S_U}|\psi(\delta,\delta)|\le \frac{11}{10}\max_{i}|\psi(u_i,u_i)|+\frac{51}{100}\sup_{\delta\in S_U}|\psi(\delta,\delta)|, \,\,\, \mb{or}\\
&& \sup_{\delta\in S_U}|\psi(\delta,\delta)|\le 3\max_i|\psi (u_i,u_i)|.
\eenrr
Applying (\r{lw1}) of Lemma \r{loh} to each $\psi(u_i,u_i)$ and taking a union bound over the $100^{m+s}$ such possibilities we obtain,
$$
P\left(\sup_{\delta\in S_U}\Big|\frac{\|X\delta\|_2^2}{n}-E\frac{\|X\delta\|_2^2}{n}\Big|\ge 3t\right) \le 2\, 100^{m+s}\, \exp\Big[-c_0n\min\big(\frac{t^2}{\si^4_x},\frac{t}{\si^2_x}\big)\Big].
$$
Again taking the union bound over all ${p\choose m}\le p^m$ possibilities of U we obtain
$$
P\left(\sup_{\delta\in \cB_1(m)}\Big|\frac{\|X\delta\|_2^2}{n}-E\frac{\|X\delta\|_2^2}{n}\Big|\ge 3t\right) \le 2\, 100^{m+s}\, p^m\exp\Big[-c_0n\min\big(\frac{t^2}{\si^4_x},\frac{t}{\si^2_x}\big)\Big].\nn
$$
Choose $t$ such that
$$
c_0\sqrt{n}t/\si^2_x=\sqrt{m\log p}+ \sqrt{(m+s)\log 100}+ \sqrt{(m+s)+\log (1/c_3)}.
$$
Then for $n$ large enough $t^2<t$ and
we obtain
\benrr
P\left(\sup_{\delta\in \cB_1(m)}\Big|\frac{\|X\delta\|_2^2}{n}-E\frac{\|X\delta\|_2^2}{n}\Big|\ge r_n\right) \le 2c_3\exp[-m-s].
\eenrr
Summing over all $m$ both sides of this bound in turn yields
\benrr
\lefteqn{
P\left(\sup_{\delta\in \cB_1(m)}\Big|\frac{\|X\delta\|_2^2}{n}-E\frac{\|X\delta\|_2^2}{n}\Big|\ge r_n,\,\,\textrm{for all}\, m\le n\right) } \hs .5in  \\
&\le & 2c_3\sum_{m=0}^{\iny}\exp[-m-s]
= 2c_3 e^{-s}/(1-1/e).\nn
\eenrr
This completes the proof of  the lemma \hfill$\Box$

\vspace{3mm}
\noi {\bf Proof of Lemma \r{r1}.} {\bf Case 1: Additive Error:} Apply Lemma \r{l2} to the sub-Gaussian matrix $Z$ to obtain w.p. at least  $1-2c_3 \exp(-s)/(1-1/e),$ for any $\delta \in \cB_1(m),$ $m\le k_n,$
\benrr
\Big|\delta'\big[n^{-1}Z'Z-\Si_z\big]\delta\Big|\le r_n(m,c_3).
\eenrr
Now substitute the relation $\Si_z=\Si_x+\Si_w$ in this inequality to obtain
\benr\lel{r1ine1}
\delta'\Si_x\delta-r_n(m,c_3)\le\delta'\Gamma^{\mb{add}}\delta\le \delta'\Si_x\delta+r_n(m,c_3).
\eenr
Notice that since $k_n\log p=o(n)$ by assumption, $r_n\to 0.$ Hence (\r{r1ine1}) and the assumption (\r{xeigen}) imply that the condition {\bf RSE$(k_n)$} holds for the matrix $\G^{\mb{add}}$ w.p. at least $1-2c_3 \exp(-s)/(1-1/e)$, for all $n$ large.

\vspace{2mm}
\noi{\bf Case 2: Missing Covariates:} Observe that for any $\delta \in \R^p$ we have
\benrr 
\Big|\delta'(\G^{\mb{miss}}-\Si_x)\delta\Big|&=&\Big|\delta'\Big(\big(n^{-1}Z'Z-\Si_z\big)\ominus M\Big)\delta\Big| \\
&\le& \frac{1}{(1-\rho_{\max})^2}\Big|\delta'\big(n^{-1}Z'Z-\Si_z\big)\delta\Big|. \nn
\eenrr
Since $Z$ is sub-Gaussian by Remark \r{zsubg}, applying Lemma \r{l2} we obtain w.p. at least $1-2c_3 \exp(-s)/(1-1/e)$ that
\benrr
\Big|\delta'(\G^{\mb{miss}}-\Si_x)\delta\Big|\le \frac{r_n(m,c_3)}{(1-\rho_{\max})^2}\,.
\eenrr
The fact $0\le \rho_{\max}<1$ and the assumption (\r{xeigen}) imply that the condition {\bf RSE$(k_n)$} holds for the matrix $\G^{\mb{miss}}$, w.p. at least $1-2c_3 \exp(-s)/(1-1/e)$, for large enough $n.$ \hfill$\Box$

\subsection{Proofs for Section 4}
\noi {\bf Proof of Theorem \r{mst1}.}
Part (ii) of this theorem follows by construction of $\h T(a_n).$ We prove part (i) separately for the two cases of additive errors and missing covariates.

\vspace{2mm}
{\noi {\bf Case 1:  Additive Error:}} It suffices to show that except on a set with asymptotic probability zero,
\benr\lel{zy}
\max|Z_{T^c}'y|< \min|Z_{T}'y|.
\eenr
Consider
\benr\lel{ini1}
\big\|\h\g-\Si_x\b_0\big\|_{\iny}&=&\big\|n^{-1}Z'y- \Si_x\b_0\big\|_{\iny}\\
&\le&\big\|n^{-1}Z'X\b_0-\Si_x\b_0\big\|_{\iny}+\big\|n^{-1}Z'
\vep\big\|_{\iny}. \nn
\eenr
Let
\benr\lel{t1}
t_1=c_0\si_{z}\si_{\vep}
\sqrt{(\log p)/n}.
\eenr
Use (\r{lw2}) of Lemma \r{loh} with $t=t_1,$ to obtain that
\benr\lel{eq1}
P\Big(n^{-1}\|Z'\vep\|_{\iny}\ge t_1\Big)&\le&
6p\,\exp\Big(-cn\min\big\{\frac{t_1^2}{\si_z^2\si_\vep^2},\frac{t_1}
{\si_z\si_{\vep}}\big\}\Big) \\
&\le &  c_1\exp\big(-c_2\log p\big). \nn
\eenr
Similarly, upon choosing $t_2= c_0\si_{z}\si_{x}\|\b_0\|_2\sqrt{(\log p)/n}$ we obtain
\benr\lel{eq2}
P\Big(n^{-1}\|Z'X\b_0-\Si_x\b_0\|\ge t_2\Big)\le c_1\exp\big(-c_2\log p\big).
\eenr
Using inequalities (\r{eq1}) and (\r{eq2}) in (\r{ini1}), we obtain w.p. at least $1-c_1\exp(-c_2\log p),$
\benrr\lel{ini2}
\big\|\h\g-\Si_x\b_0\big\|_{\iny}\le t_1+t_2,
\eenrr
for all sufficiently large $n$. The proof of (\r{zy}) is now completed upon combining this bound 
with assumption {\bf (A3i)}.
This concludes the proof for additive errors.

\vspace{2mm}
\noi {\bf Case 2: Missing Covariates:} Here, it suffices to show that except on a set with asymptotic probability zero we have,
\benr\lel{zym}
\max|Z_{T^c}'y \ominus ({\bf 1}-{\boldsymbol \rho})|< \min|Z_{T}'y\ominus ({\bf 1}-{\boldsymbol \rho})|.
\eenr

For this purpose consider
\benr\lel{infi}
\lefteqn{
\big\|\h\g-\Si_x\b_0\big\|_{\iny}  } \\
&=&\Big\|n^{-1}Z'y\ominus ({\bf {1}}-{\boldsymbol\rho})- \Si_x\b_0\|_{\iny}\Big\|_{\iny} \nn\\
&\le& \frac{1}{1-\rho_{\max}}\Big(\big\|
n^{-1}Z'X\b_0-{\mb{cov}}(z_i,x_i)\b_0\big\|_{\iny}+\|n^{-1}Z'\vep\|\Big). \nn
\eenr
Recall as stated in Remark \r{zsubg}, $Z$ is a sub-Gaussian matrix with parameter $\si_x.$ Hence, argue as for (\r{eq1}) and (\r{eq2}), with $t_1$ as in (\r{t1}) and $t_2=c_0\si_{x}^2\|\b_0\|_2\sqrt{(\log p)/n},$ to obtain that
\benrr
&& P\Big(n^{-1}\|Z'\vep\|_{\iny}\ge t_1\Big)\le c_1\exp(-c_2\log p),\\
&& P\Big(n^{-1}\|Z'X\b_0-\Si_x\b_0\|\ge t_2\Big)\le c_1\exp(-c_2\log p).
\eenrr
These bounds together with the inequality (\r{infi}) imply that, w.p. at least $1-c_1\exp(-c_2\log p)$,
\benrr\lel{eqin}
\|\h\g-\Si_x\b_0\|_{\iny}\le \frac{1}{1-\rho_{\max}}[t_1+t_2],
\eenrr
for all sufficiently large $n$. The claim (\r{zym}) now follows from this bound,
assumption {\bf (A3ii)} and the assumption {\bf (A2)} that ensures  $0\le\rho_{\max}<1$.  \hfill $\Box$

\vspace{2mm}
\noi {\bf Proof of Theorem \r{mst2}.} Recall that this theorem pertains only to the case of missing covariates. Part (i) of this theorem is a consequence of Theorem 1 of Loh and Wainwright (2012), who show that under the assumed conditions, $\|\h\b-\b\|_2\le \la_n c_0\sqrt{s}/\al_1$, w.p. at least $1-c_1\exp(-c_2\log p)$, where $\al_1$ is as in assumption {\bf RE}. This result together with assumption {\bf (A4)} implies that $T\subseteq \h T$, with the same probability, for all sufficiently large $n$.

To prove part (ii) notice that by the first order optimality conditions $\Big(\Gamma\h\b-\h\g\Big)_{\h T}=\la_n,$ 
\benr\lel{meq1}
\sqrt{\mb{card}(\h T)}\la_n&\le& \Big\|\Big(\Gamma\h\b -\h\g\Big)_{\h T}\Big\|_2 \\
&\le& \sqrt{\mb{card}(\h
T)}\left(\big\|\G\b_0-\h\g\big\|_{\iny}\right)+\Big\|\big(\Gamma(\h\b-\b_0)\big)_{\h T}\Big\|_2 \nn \\
&=& (I)+(II), \qquad \mb{(say)}. \nn
\eenr

Let $v$ be a unit vector such that $\h\b-\b_0=\|\h\b-\b_0\|_2v$ and note that $\|v_{T^c}\|_0\le \h m.$ Consider the term (II) on the r.h.s of (\r{meq1}).
\benr\lel{meq2}
\lefteqn{
\Big\|\big(\Gamma (\h\b-\b_0)\big)_{\h T}\Big\|_2 } \\
&\le& \|\h\b-\b_0\|_2\sup_{\|\delta_{T^c}\|_{0}\le \h m, \|\delta\|_2\le 1} \big|\delta'\Gamma v\big|\nn\\
&\le& 3\|\h\b-\b_0\|_2 \sup_{\|\delta_{T^c}\|_{0}\le \h m, \|\delta\|_2\le 1} \big|\delta'\Gamma \delta\big|\le 3\phi(\h m)\|\h\b-\b_0\|_2.  \nn
\eenr
Now consider term (I) of (\r{meq1}) for the case of missing covariates.
\benr
\big\|\G\b_0-\h\g\big\|_{\iny}&\le& \|\G\b_0-\Si_x\b_0\|_{\iny} +\|\h\g-\Si_{x}\b_0\|_{\iny}\nn\\
&\le& \big\|\Big(\big(n^{-1}Z'Z-\Si_z\big)\ominus M\Big)\b_0\big\|_{\iny}\nn\\
&&+\frac{1}{1-\rho_{max}}\Big(\big\|n^{-1}Z'X\b_0-\Si_x\b_0\big\|_{\iny}+\big\|n^{-1}Z'\vep\big\|_{\iny}\Big)\nn\\
&\le& c_0 \frac{\si_x}{1-\rho_{\max}}\big(\si_{\vep}+\frac{\si_x}{1-\rho_{\max}}\big)\|\b_0\|_2\sqrt{\frac{\log p}{n}}\le \la_n/2. \nn
\eenr
Here the second inequality follows by basic algebra. The third inequality holds w.p. at least $1-c_1\exp(-c_2\log p)$ which follows by applying (\r{lw2}) of Lemma \r{loh} separately on each of the three terms. The final inequality follows from the choice of $\la_n$ under the missing covariate case.  Combine this result with (\r{meq2}) and (\r{meq1}) to obtain
\benrr
\sqrt{\mb{card}(\h T)}\la_n[1-2^{-1}]\le 3\phi(\h m)\|\h\b-\b_0\|_2\,.
\eenrr
On the other hand by part (i), $\|\h\b-\b_0\|_2\le \la_n c_0\sqrt{s}/\al_1$, w.p. at least $1-c_1\exp(-c_2\log p),$ for all sufficiently large $n$. These facts  together with the fact $\h m\le \mb{card}(\h T)$ readily imply
$ 
\sqrt{\h m}\le 4\phi (\h m) c_0\sqrt{s}/\al_1,
$ 
w.p. at least $1-c_1\exp(-c_2\log p),$ for all sufficiently large $n$.
This
completes the proof of the Theorem \r{mst2}.\hfill $\Box$

\subsection{Proofs for Section 5} The proofs for this section shall require the following series of three lemmas. To proceed further we need to define
\benr\lel{r}
r=\begin{cases}
      \si_z\si_{\vep}\,\,; & \mb{for additive errors} \\
     \si_x\si_{\vep}\,\,; & \mb{for missing covariates}
            .\end{cases}
\eenr
The structure of the proof of the following two lemma's is similar to the proof of Lemma \r{l2}. All three results provide uniform bounds that hold in probability on different random quantities, for all sufficiently large $n$.

\begin{lem}\lel{el3}
Let $r$ be as in (\r{r}). Then, uniformly over all $m\le n$ and for any $c_3\in (0,1)$ and some universal constant $D$,
\benrr
\sup_{\delta\in\cB(m),\,\|\del\|_2>0}\Big|\frac{\delta'Z'\vep}{n\|\delta\|_2}\Big|\le c_0\frac{3r}{2}e_n(m,c_3),
\eenrr
w.p. at least $1-2c_3 e^{-s}/(1-1/e),$ for all sufficiently large $n$.
\end{lem}

\noi {\bf Proof of Lemma \r{el3}}
 For every $U\subseteq\{1,...,p\}$ with $\mb{card}(U-T)\le m,$ define $S_U:=\{\delta\in\R^p;\, \|\delta\|_2\le 1;\, \textrm{supp}(\delta)\subseteq U\}$ and note that $\cB_1(m)= \cup_{card(U-T)\le m}S_U.$ Let $\cA=\{u_1,...,u_k\}$ be a $1/3$ cover of a fixed $S_U,$ i.e., $\forall \delta\in S_U$ $\exists\,\,u_i\in\cA$ such that $\|\delta-u_i\|_2\le 1/3.$ It is known from Ledoux and Talagrand (1991) or Loh and Wainwright (2012) (Supplementary materials, pg.17), that we can construct $\cA$ such that $card(\cA)\le 9^{(m+s)}.$ Then by elementary algebra,
\benr\lel{l3eq1}
n^{-1}\delta'Z'\vep= n^{-1}u_i'Z'\vep + n^{-1}(\delta-u_i)'Z'\vep
\eenr
By construction of $\cA,$ $3(\delta-u_i)\in S_U$ and using (\r{l3eq1}) we obtain,
\benrr
\sup_{\delta\in S_U}\big|n^{-1}\delta'Z'\vep\big|\le \max_{i} \big|n^{-1}u_i'Z'\vep\big| + \sup_{\delta\in S_U}\big|\frac{1}{3n}\delta'Z'\vep\big|.
\eenrr
Hence $\sup_{\delta\in S_U}\big|n^{-1}\delta'Z'\vep\big|\le \max_{i} \big|\frac{3}{2n}u_i'Z'\vep\big|.$ Now applying Lemma \r{loh}, $9^{m+s}$ times, once for each $n^{-1}u_i'Z'\vep$ and taking a union bound over all such possibilities we obtain,
\benr
P\left(\sup_{\delta\in S_U}\Big|n^{-1}\delta'Z'\vep\Big|\ge \frac{3t}{2}\right) \le 2\cdotp9^{m+s}\cdotp \exp\Big[-\frac{1}{c_0}n\min\big(\frac{t^2}{(\si_z\si_{\vep})^2},\frac{t}{\si_z\si_{\vep}}\big)\Big].\nn
\eenr
Again taking the union bound over all ${p\choose m}\le p^m$ possibilities of U we obtain
\benr
P\left(\sup_{\delta\in \cB_1(m)}\Big|n^{-1}\delta'Z'\vep\Big|\ge \frac{3t}{2}\right) \le 2\cdotp9^{m+s}\cdot p^m\cdotp \exp\Big[-\frac{1}{c_0}n\min\big(\frac{t^2}{(\si_z\si_{\vep})^2},\frac{t}{\si_z\si_{\vep}}\big)\Big].\nn
\eenr
Choose $t=re_n(m,c_3)$ to obtain,
\benr
P\left(\sup_{\delta\in \cB_1(m)}\Big|n^{-1}\delta'Z'\vep\Big|\ge \frac{3t}{2}\right) \le  2c_3\exp[-m-s].\nn
\eenr
and thus
\benr
P\left(\sup_{\delta\in \cB_1(m)}\Big|n^{-1}\delta'Z'\vep\Big|\ge \frac{3t}{2},\,\,\,\textnormal{for any}\,\,m\right) &\le & 2c_3\sum_{m=0}^{\iny}\exp[-m-s]\nn\\
&=& 2c_3 \exp(-s)/(1-1/e),\nn
\eenr
thereby completing the proof of the lemma. \hfill$\Box$

\begin{lem}\lel{el4}
{Let $r$ be as in (\r{r}).}
Then uniformly over all $m\le n$ and for any $c_3\in (0,1)$ and some universal constant $D$ we have,
\benr
\sup_{\delta\in\cB(m),\,\|\del\|_2>0}\frac{1}{\|\delta\|_2}\Big|\delta'\big[\frac{Z'W}{n}-\Si_x\big]\b_0\Big|\le c_0\frac{3r}{2}\|\b_0\|_2e_n(m,c_3),
\eenr
w.p. at least $1-2c_3 e^{-s}/(1-1/e),$ for all sufficiently large $n$.
\end{lem}

\noi {\bf Proof of Lemma \r{el4}.}
\vspace{2mm} Following the same idea as in the proof of Lemma \r{el3}, construct an $1/3$ cover $\cA$ of $S_U$ for each $U$ and let $\Big|\delta'\big[\frac{Z'W}{n}-\Si_w\big]\b_0\Big|=\psi_{zw}(\delta,\b_0).$ Then
\benrr
\psi_{zw}(\delta,\beta^0)=\psi_{zw}(u_i,\b_0)+\psi_{zw}(\del-u_i,\b_0).
\eenrr
This in turn implies that
\benrr
\sup_{\del\in S_U}\big|\psi_{zw}(\delta,\b_0)\big|& \le& \max_{i}\big|\psi_{zw}(u_i,\b_0)\big|+ \frac{1}{3}\sup_{\delta\in S_U} \big|\psi_{zw}(\delta,\b_0)\big|, \,\,\, \mb{or} \\
\sup_{\del\in S_U}\big|\psi_{zw}(\delta,\b_0)\big| &\le& \frac{3}{2}\max_{i}\big|\psi_{zw}(u_i,\b_0)\big|.
\eenrr
Now, use Lemma \r{loh} to obtain
$$
P\left(\sup_{\delta\in\cB_1}\big|\psi_{zw}(\delta,\b_0)\big|>\frac{3t}{2}\right)\le 2\, 9^{m+s}\, p^m \, \exp\Big[-c_0 n \min \big(\frac{t^2}{(\si_z\si_w)^2},\frac{t}{\si_z\si_w}\big)\Big].
$$
Choosing $t=c_0\frac{3r}{2}e_n(m,c_3)$ we obtain,
\benr\lel{el1r1}
P\left(\sup_{\delta\in\cB_1(m)}\big|\psi_{zw}(\delta,\b_0)\big|>\frac{3t}{2},\,\,\textrm{for any}\, m\le n\right) &\le & 2c_3\sum_{m=0}^{\iny}\exp[-m-s]\nn\\
&=& 2c_3 e^{-s}/(1-1/e).
\eenr
Thus we obtain uniformly over any $\delta\in \cB_1(m)$ and $m\le n,$
\benr
\Big|\delta'\big[\frac{Z'W}{n}-\mb{cov}(z_i,w_i)\big]\b_0\Big|\le c_0\frac{3r}{2}\|\b_0\|_2e_n(m,c_3),
\eenr
w.p. at least $1-2c_3 e^{-s}/(1-1/e),$ for all sufficiently large $n$, thereby completing the proof. \hfill $\Box$

\vspace{2mm}
\noi {\bf Proof of Lemma \r{udcc}.}
{\bf Case 1: Additive error:} We have
\benr\lel{edcin1}
\Big|\delta'(\G^{\mb{add}}\b_0-\h\g^{\mb{add}})\Big|\le \Big|\frac{\delta'Z'\vep}{n}\Big|+\Big|\delta'
\big[\frac{Z'W}{n}-\Si_w\big]\b_0\Big|.
\eenr
Apply Lemmas \r{el3} and  \r{el4} to the two terms on the r.h.s.\,\,of this bound and substitute back in (\r{edcin1}) to obtain the desired result.

\vspace{1mm}
\noi{\bf Case 2: Missing Covariates:}
Proceeding as in {\bf Case 2} of the proof of Theorem \r{mst2}, we obtain
\benrr
\big|\delta'\G^{\mb{miss}}\b_0-\h\g^{\mb{miss}}\big|&\le& \big|\delta'\big[\G^{\mb{miss}}-\Si_x\big]\b_0\big| +\big|\delta'\big(\h\g^{\mb{miss}}-\Si_{x}\b_0\big)\big|\nn\\
&\le& \frac{1}{(1-\rho_{max})^2}\Big|\delta'\big(n^{-1}Z'Z-\Si_z\big)\b_0\big|\nn\\
&&+\frac{1}{1-\rho_{max}}\left(\Big|\delta'\big(\frac{Z'X\b_0}{n}-\Si_x\b_0\big)\Big|
+\big|\delta'n^{-1}Z'\vep\big|\right).
\eenrr
The claim of the lemma again follows by applying Lemmas \r{el3} and \r{el4} to the last expression.
\hfill$\Box$

\begin{lem}\lel{el1}
Let
\benr
r=\begin{cases}
      \si_z(\si_w+\si_{\vep})\hs 0.6in\,; & \mb{for additive error} \\
      \frac{\si_x}{1-\rho_{\max}}\big(\si_{\vep}+\frac{\si_x}{1-\rho_{\max}}\big)\,\,; & \mb{for missing covariates},
       \end{cases}
\eenr
Then uniformly over all $\delta\in\cB(m),$ $m\le n$ and for any $c_3\in (0,1)$ and some universal constant D we have,

\benr
\big|\h Q_n(\b_0+\del)-\h Q_n(\b_0)-\del'\G\del/2\big|\le c_0\|\delta\|_2\|\b_0\|_2re_n(m,c_3),\nn
\eenr
w.p. at least $1-6c_3 e^{-s}/(1-1/e).$
\end{lem}


\noi {\bf Proof of Lemma \r{el1}.} This lemma is a straightforward consequence of Lemma \r{udcc}. Using the definition of $\h Q_n(\cdotp)$ we obtain,
\benr\lel{el1in2}
\big|\h Q(\b_0+\del)-\h Q(\b_0)-\delta'\G\delta/2\big|&=&\big|\delta'(\G\b_0-\h\g)\big|\nn
\eenr
Lemma \r{udcc} applied to the r.h.s.\,\,of this equation yields the desired result.  \hfill$\Box$

\vspace{2mm}
\noi {\bf Proof of Theorem \r{t2}.} Let $\tilde \delta=\tilde \b-\b_0,$ then by Lemma \r{el1},  w.p. at least $1-6c_3\exp(-s)/(1-1/e),$
\benrr
\big|\h Q(\b_0+\tilde\del)-\h Q(\b_0)-\tilde\delta'\G\tilde\del/2\big|\le c_0r\|\tilde\del\|_2\|\b_0\|_2e_n(m,c_3).
\eenrr
Also by the model selection step, $T\subseteq \h T$ w.p. at least $1-c_1\exp(-c_2\log p).$ Hence, on this set, by the definition of the second step estimator, $\h Q(\tilde\b)-\h Q(\b_0)\le 0.$  This in turn implies that w.p. at least $1-6c_3\exp(-s)/(1-1/e)-c_1\exp(-c_2\log p),$
\benr\lel{et2in1}
-\tilde\delta'\G\tilde\del/2\ge -c_0r\|\tilde\del\|_2\|\b_0\|_2e_n(m,c_3).
\eenr
An application of condition RSE$(\h m)$ in the inequality (\r{et2in1}) yields
\benr
\|\tilde\delta\|_2\le\frac{c_0r}{\ka_x(\h m)} \|\b_0\|_2e_n(m,c_3),
\eenr
w.p. at least $1-6c_3\exp(-s)/(1-1/e)-c_1\exp(-c_2\log p).$
This completes the proof of this Theorem.\hfill$\Box$

The \noi {\bf Proof of Theorem \r{gmc}} shall rely on the following two results. First is Lemma 6 of the supplement of Loh and Wainwright (2012), which is restated below for the convenience of the reader.


\begin{lem}\lel{gmint} For each $1\le j\le p,$
\benrr
&& \frac{1}{\la_{\max}(\Si)}\le|d_j|\le \frac{1}{\la_{\min}(\Si)}\quad \mb{and}\quad \|\theta^j\|_2\le \la_{\max}(\Si)/\la_{\min}(\Si)
\eenrr
\end{lem}


\begin{lem}\lel{gmml} Under the conditions of Theorem \r{gmc} the following hold.
\benrr
&&\mb{(i)}\,\,\,|\h d_j- d_j|\le c_0C_2e_n(a_n-s,c_3), \nn\\
&& \mb{(ii)}\,\,\|\tilde \Theta_{\cdotp j}-\Theta_{\cdotp j}\|_2 \le c_0 \Big(C_2^2+\frac{C_1^2}{\la_{\min}^2(\Si)}+\frac{\la_{\max}(\Si)}
{\la_{\min}(\Si)}C_2\Big)^{1/2} e_n(a_n-s, c_3),
\eenrr
for all $1\le j\le p$, w.p. converging to $1.$
\end{lem}

\noi {\bf Proof of Lemma \r{gmml}}
Let $\h m=a_n-s$ and observe that in view of Theorem \r{mst1} and Theorem \r{t2} we have for all $1\le j\le p,$
\benr
\|\h\theta^j-\theta^j\|_2\le \frac{1}{\ka(\h m)}c_0r\|\theta^j\|_2e_n(\h m,c_3):=c_0C_1e_n(\h m,c_3),
\eenr
w.p.  converging to $1.$ Also, note that by the additional parameter space restriction in the construction of (\r{gest}),  $\|\h\theta^{j}\|_1\le b_0\sqrt{s}.$ Consider
\benr\lel{gmli1}
|\h d_j^{-1}-d_j^{-1}|&=&\Big|\big(\h\Si_{jj}-\h\Si_{j,-j}\h\theta^{j}\big)-\big(\Si_{jj}-\Si_{j,-j}\theta^{j}\big)\Big|\nn\\
&\le& \big|\h\Si_{jj}-\Si_{jj}\big|+\big|\h\Si_{j,-j}\h\theta^j-\Si_{j,-j}\theta^j\big|:= (I)+(II), \,\, \mb{(say)}.
\eenr
By assumption we have that $(I)\le c_0\si_x\sqrt{{\log p}/{n}}$. Now consider the term  (II) on the r.h.s of (\r{gmli1}),
\benr
(II)&\le& \Big|\big(\h\Si_{j,-j}-\Si_{j,-j}\big)\h\theta^j\Big|+\Big|\Si_{j,-j}\big(\h\theta^j-\theta^j\big)\Big|\nn\\
&\le& \|\h\Si-\Si\|_{\iny}\|\h\theta^j\|_1+\|\Si_{j,-j}\|_2\|\h\theta^j-\theta^j\|_2\nn\\
&\le& c_0\Big(b_0\si_x+\frac{\la_{\max}}{\la_{\min}}r\|\theta^j\|_2\Big)e_n(\h m,c_3).
\eenr
Combining the bounds for terms (I) and (II) we obtain for all $1\le j\le p,$
\benrr
|\h d_j^{-1}-d_j^{-1}|\le c_0\Big(b_0\si_x+\frac{\la_{\max}}{\la_{\min}}r\|\theta^j\|_2\Big)e_n(\h m,c_3).
\eenrr
Thus applying Lemma \r{gmint} we obtain,
\benrr
\Big|\frac{d_j}{\h d_j}-1\Big|\le |d_j||\h d_j^{-1}-d_j^{-1}|\le c_0\frac{1}{\la_{\min}}\Big(b_0\si_x+\frac{\la_{\max}}{\la_{\min}}r\|\theta^j\|_2\Big)e_n(\h m,c_3).\nn
\eenrr
This in turn implies that $|\h d_j|\le 2|d_j|$ for $n$ sufficiently large, and hence
\benrr
|\h d_j-d_j|\le |\h d_j|\Big|\frac{d_j}{\h d_j}-1\Big|&\le& c_0\frac{1}{\la_{\min}^2}\Big(b_0\si_x+\frac{\la_{\max}}{\la_{\min}}r\|\theta^j\|_2\Big)
e_n(\h m,c_3),\nn\\
&:=&c_0C_2e_n(\h m,c_3),
\eenrr
for $n$ sufficiently large. This proves part (i) of this lemma. To prove (ii) consider,
\benrr
\|\tilde \Theta_{\cdotp j}-\Theta_{\cdotp j}\|_2^2&=&|\h d_j- d_j|^2 + \|\h d_j\h\theta^j-d_j\theta^j\|_2^2\nn\\
&\le& |\h d_j- d_j|^2 + 2|d_j|^2\|\h\theta^j-\theta^j\|_2^2+2|\h d_j-d_j|^2\|\h\theta^j\|_2^2\nn\\
&\le& c_0^2 \Big(C_2^2+\frac{C_1^2}{\la_{\min}^2(\Si)}+\frac{\la_{\max}}{\la_{\min}}C_2\Big) e_n^2(\h m, c_3)\nn
\eenrr
This completes the proof of the lemma. \hfill $\Box$

\noi {\bf Proof of Theorem \r{gmc}}
This proof is a direct consequence of Lemma \r{gmml} by observing that
\benrr
&&\|\h\Theta-\Theta\|_2^2\le 2\|\h\Theta-\tilde\Theta\|_2^2+ 2\|\tilde\Theta-\Theta\|_2^2\le 4\max_{j}\|\tilde\Theta_{\cdotp j}-\Theta_{\cdotp j}\|_2^2. \hs .6in  \Box
\eenrr

\section*{References}
\begin{enumerate}
\item Agarwal, A., Neghban, S. and Wainwright, M.J. (2012). Fast Global Convergence of gradient methods for High Dimensional Statistical Recovery.  {\it Ann.  Statist.}  {\bf 40}, 2452--2482.

\item Belloni, A., and Chernozhukov, V. (2013). Least Squares After Model Selection in High Dimensional Sparse Models,  {\it Bernoulli}. {\bf 19}, 521--547.

\item Bickel, P., Ritov, Y. and Tsybakov, A. (2009). Simultaneous Analysis of Lasso and Dantzig Selector, {\it Ann. Statist.}, {\bf 37}, 1705--1732.

\item Bickel, P., Levina, E. (2008). Covariance Regularization by Thresholding, {\it Ann. Statist.}, {\bf 36}, 2577--2604.

\item B\"{u}hlmann, P. and van de Geer, S. (2011). {\it Statistics for High Dimensional Data}. Springer-Verlag, Berlin Heidelberg.

\item B\"{u}hlmann, P., Kalisch, M., and Maathuis, M. H. (2009). Variable Selection in High Dimensional Linear Models; Partially Faithful Distributions and the PC-Simple Algorithm.  {\it Biometrika}. {\bf{97}}, 261-278.

\item Carroll, R.J., Ruppert, D., Stefanski, L.A. and Crainiceanu, C. (2006).
{\it Measurement Error in Nonlinear Models: A Modern Perspective}. Chapman \& Hall, New York.

\item Fan, J. and Lv, J. (2008). Sure Independence Screening for Ultrahigh Dimensional Feature Space . {\it J.R. Stat. Soc. Ser. B Stat. Methodol.} {\bf 70} 849--911.

\item Friedman, J., Hastie, T., Simon, N., Tibshirani, R. (2010). Regularization Paths for Generalized Linear Models via Coordinate Descent.  {\it J. Statist. Software}, {\bf 33}, 1--22.

\item Friedman, J., Hastie, T., Simon, N., Tibshirani, R. (2008). Sparse Inverse Covariance Estimation with the Graphical Lasso. {\it Biostatistics}, {\bf9}, 432--441.

\item Fuller, W.A. (1987). {\it Measurement Error Models}. Wiley \& Sons, Inc. New York.

\item Genovese, C., Jin, J., Wasserman, L., Yao, Z. (2012). A Comparison of the Lasso and Marginal Regression, {\it J. of Mach. Learn. Res.},  {\bf 13}, 2107--2143.

\item Kaul, A. and Koul, H. (2015). Weighted $\ell_1$-Penalized Corrected Quantile Regression for High Dimensional Measurement Error Models. {\it J. Mult. Analysis.}, {\bf 140}, 72--91.

\item Ledoux, M. and Talagrand, M. (1991). Probability in Banach Spaces: Isoperimetry and Processes. Springer-Verlag, New York.

\item Liang, H. and Li, R. (2009). Variable Selection for Partially Linear Models with Measurement Errors. {\it J. Amer. Statist. Assoc.}, {\bf 104}, 234--248.

\item Loh, P., and Wainwright, M.J. (2012). High-dimensional regression with noisy and missing data: Provable guarantees with non-convexity. {\it Annals of Statistics} , {\bf{40}}, 1637--1664.

\item Meinhausen, N. and B\'uhlmann, P. (2006) High Dimensional graphs and Variable Selection with Lasso. {\it Annals of Statistics}, {\bf 34}, 1436--1462.

\item Pang, H., Liu, H., and Vanderbei, R. (2014). The fastclime Package for Linear Programming and Large-Scale Precision Matrix Estimation in R.
    {\it J. Mach. Learn. Res.}, {\bf{15}} 489-493.

\item Rosenbaum, M. and Tsybakov, A.B. (2010). Sparse recovery under matrix uncertainty. {\it Annals of Statistics}, {\bf{38}} 2620--2651.

\item Rosenbaum, M. and Tsybakov, A.B. (2011). Improved matrix uncertainty selector,
{\it Technical Report}.  Available at http://arxiv.org/abs/1112.4413.

\item S\o rensen, \O., Frigessi, A., and Thoresen, M. (2014). Covariate Selection in High-Dimensional Generalized Linear Models With Measurement Error.
    Available at http://arxiv.org/abs/1407.1070.

\item S\o rensen, \O., Frigessi, A., and Thoresen, M. (2015). Measurement error in lasso: impact and likelihood bias correction. {\it Statist. Sinica}, {\bf 25(2)}, 809--829.

\item van der Waart, A. W. and Wellner, J. A. (1996). {\it Weak Convergence and Empirical Processes}.  Springer, New York.

\item Tibshirani, R.J. (2013). The lasso problem and uniqueness. {\it Electron. J.  Stat.}, {\bf{7}}, 1456--1490.

\item Vershynin. (2012). Introduction to the Non-Asymptotic Analysis of Random Matrices. {\it Chapter 5 of Compressed Sensing: Theory and Applications}. Cambridge University Press.

\item Yuan, M. (2010). High Dimensional Inverse Covariance Matrix Estimation via Sparse Linear Programming, {\it J. of Mach. Learn. Res.}, {\bf{11}} 2261--2286.
\end{enumerate}

%
%
%
%
%
%
%

\edt